\documentclass[]{emulateapj}
\usepackage{epsfig, natbib, graphicx, color}

\input epsf 

\usepackage{color}

\newcommand{\Mpc}{\mbox{Mpc}}

\newcommand{\msun}{M_\odot}

\newcommand{\fgas}{f_{\rm gas}}

\newcommand{\avg}[1]{\left\langle #1 \right\rangle}

\newcommand{\Mf}{M_{500}}

\newcommand{\be}{\begin{equation}}
\newcommand{\ee}{\end{equation}}
\newcommand{\bea}{\begin{eqnarray}}
\newcommand{\eea}{\end{eqnarray}}

\newcommand{\keV}{\mbox{keV}}
\newcommand{\Ysz}{Y_{\rm SZ}}
\newcommand{\Yx}{Y_{\rm X}}
\newcommand{\Tx}{T_{\rm X}}
\newcommand{\Lx}{L_{\rm X}}

\newcommand{\rosat}{{\it ROSAT}}
\newcommand{\chandra}{{\it Chandra}}
\newcommand{\XMM}{{\it XMM-Newton}}
\newcommand{\xmm}{{\it XMM-Newton}}

\newcommand{\maxBCG}{{\it maxBCG}}

\newcommand{\rhogas}{\rho_{\rm gas}}
\newcommand{\Mgas}{M_{\rm gas}}
\newcommand{\Mtot}{M_{500}}
\newcommand{\Rf}{R_{500}}

\shortauthors{Rozo et al.}
\shorttitle{A Comparative Study of Local Galaxy Clusters:  I. Derived X-ray Observables}

\begin{document}
\title{A Comparative Study of Local Galaxy Clusters:  I. Derived X-ray Observables}
\author{Eduardo Rozo\altaffilmark{1,2}, Eli S. Rykoff\altaffilmark{3,4}, James G. Bartlett\altaffilmark{5,6}, August Evrard\altaffilmark{7}}
\altaffiltext{1}{Einstein Fellow, Department of Astronomy \& Astrophysics, The University of Chicago, Chicago, IL 60637.}
\altaffiltext{2}{Kavli Institute for Cosmological Physics, Chicago, IL 60637.}
\altaffiltext{3}{SLAC National Accelerator Laboratory, Menlo Park, CA 94025.}
\altaffiltext{4}{Lawrence Berkeley National Laboratory, Berkeley, CA 94720.}
\altaffiltext{5}{APC, AstroParticule et Cosmologie, Universit\'e Paris Diderot, CNRS/IN2P3, CEA/lrfu, Observatoire de Paris, Sorbonne Paris
Cit\'e, 10, rue Alice Domon et L\'eonie Duquet, Paris Cedex 13, France.}
\altaffiltext{6}{Jet Propulsion Laboratory, California Institute of Technology, 4800 Oak Grove Drive, Pasadena, CA, U.S.A.}
\altaffiltext{7}{Departments of Physics and Astronomy and Michigan Center for Theoretical Physics, University of Michigan, Ann Arbor, MI 48109.}

\begin{abstract}
We examine systematic differences in the derived X-ray properties of galaxy clusters as reported by three different groups:
\citet{vikhlininetal09}, \citet{mantzetal10b}, and \citet{planck11_local}.  The sample overlap between any two pairs of works
ranges between 16 to 28 galaxy clusters in common. 
We find systematic differences in most reported properties, including the total cluster mass, $\Mf$.   
The most extreme case is an average $45\% \pm 5\%$ difference in cluster mass between the \citet{planck11_local} and \citet{mantzetal10b}, 
for clusters at $z > 0.13$ (averaged over 16 clusters).  
These mass differences induce differences in cluster observables defined within an $\Rf$ aperture.
After accounting for aperture differences, we find very good agreement in gas mass estimates between the different groups.  
However, the soft-band X-ray 
luminosity, $\Lx$, core-excised spectroscopic temperature, $\Tx$, and gas thermal energy, $\Yx=\Mgas \Tx$ display mean 
differences at the 5\%-15\% level.  We also find that the low ($z\leq 0.13$) and high ($z\geq 0.13$) galaxy cluster samples in
\citet{planck11_local} appear to be systematically different: the $\Ysz/Y_X$ ratio for these two sub-samples is
$\ln (\Ysz/Y_X) = -0.06\pm 0.04$ and $\ln (\Ysz/Y_X) = 0.08 \pm 0.04$ for the low and high redshift sub-samples respectively.
\end{abstract}

\section{Introduction}

X-rays studies of galaxy clusters are an established
method for investigating the physics of the intra-cluster medium 
through cluster scaling relations \citep[\hbox{\sl e.g.,}][]{prattetal09,vikhlininetal09b,arnaudetal10,
mantzetal10b}, as well as for constraining 
cosmology with cluster abundances \citep[\hbox{\sl e.g., }][]{henryetal09,vikhlininetal09,
mantzetal10a}.  With cluster samples of increasing size and improving quality 
approaching --- e.g. \citep[XCS][]{lloyd-daviesetal11}, XXL\footnote{http://irfu.cea.fr/en/Phocea/Vie\_des\_labos/Ast/ast\_technique. php?id\_ast=3015},
and {\it eRosita} \citep{pillepichetal11} --- accurately quantifying the level of systematic uncertainties in data analysis is  becoming a significant issue.  
Instrumental uncertainty has long been recognized as a source of X-ray modeling error, and the relative cross-calibration of the \chandra\ and \xmm\ observatories has been explored in a variety of works, most recently \citet{nevalainenetal10}
and \citet{tsujimotoetal11}.  These studies find $\approx 10\%$ systematic differences in the effective area of the two instruments \citep[see also][]{snowden02}, as well as differences of $\approx 10\%-15\%$ in temperature estimates.

Differences in instrumental calibration can be further compounded by methodological differences in the data analysis
between different groups.  For instance, it has been suggested that the choice of parameterization of the intra-cluster medium
plays an important role in the derived cluster scaling relations \citep{mantzallen11}.  
Theoretically, these {\sl instrumental} and {\sl methodological} sources of error should be independent.  In practice, however,
the degree of independence is compromised because any given research group often relies heavily on a single method applied to a specific instrument.  Consequently, disentangling these two sources of possible systematic errors is difficult.

Relative to instrumental calibration, methodological sources of error have received comparatively little attention.  
Codes to compute plasma emission, principally SPEX/MEKAL and APEC, produce similar continuum, but differ in expectations for line emission even at relatively low spectral resolution\footnote{http://www.atomdb.org/issues.php}.  
In addition, deriving X-ray observables from raw photon counts requires that the X-ray background and hydrogen column density to the cluster be estimated, points sources be masked, 
intra-cluster gas metallicities fit for or assumed, and data cuts applied (for instance, with respect to the energy range used to fit
for the spectral X-ray temperature).   All of these procedures can lead to systematic differences of varying degree.

In this work, we take a pragmatic approach to estimating the level of systematic differences in derived X-ray cluster properties.  With the maturation of the \xmm\ and \chandra\ observatories, the likelihood of a nearby cluster having been sensitively imaged by both observatories has increased to the point where several dozen such systems now exist.   We examine the differences in pairs of such measurements performed by different groups, focusing on three samples used for recent cosmological studies:  the Vikhlinin sample, composed of the systems in \citet{vikhlininetal06} and \citet{vikhlininetal09}, henceforth referred to
as the V09 sample;  the Mantz sample, composed of systems in \citet{mantzetal10b}, henceforth referred to as the M10 sample;  
and the Planck--XMM sample, composed of systems in \citet{planck11_local}, henceforth referred to as the P11-LS sample.   
The ``-LS'' signifies that this is the local scaling relations paper from the Planck Early Results series of papers.
We will also briefly consider clusters in the \citet{pointecouteauetal05} sample.  Published properties include the soft-band X-ray luminosity, $L_x$, intracluster gas mass, $\Mgas$, gas temperature, $\Tx$, gas thermal energy, $\Yx$, and the
hydrostatic mass $\Mf$, defined below.   We summarize the observables under consideration in Table \ref{tab:observables}.


\begin{deluxetable*}{ll}
\tablewidth{0pt}
\tablecaption{Cluster Observables}
\tablehead{Observable &  Definition }
\startdata
$\Lx$ & soft, $[0.1,2.4]\ \keV$, X-ray luminosity within a cylindrical aperture of radius, $\Rf$. \\
$\Mgas$ & Gas mass within a sphere of radius, $\Rf$. \\
$\Tx$ & Core excised spectroscopic X-ray temperature with cylindrical annulus, $R\in[0.15,1]\Rf$.\footnote{Note that different works use slightly different energy ranges when fitting cluster spectra, a potential source of variation in $\Tx$ that we do not account for.} \\
$\Yx$ & Gas thermal energy as derived from the product, $\Mgas \Tx$. \\
$\Mtot$ & Total mass, calibrated using hydrostatic mass estimates.
\enddata
\label{tab:observables}
\end{deluxetable*}


Our comparisons are based on a total of 16 clusters shared between the Vikhlinin and Mantz samples, 
23 common clusters between the Vikhlinin and Planck samples, and
28 common clusters between the Planck and Mantz samples.  We note that of these last 28 systems,
only 12 of them have temperatures estimated by \citet{mantzetal10b}.  
Consequently, the $\Tx$ and $\Yx$ comparison for this pair of samples 
is limited to these systems only.
We have restricted ourselves to low-redshift ($z\leq 0.3$) systems in this 
work.
We also note that there are 6 clusters in common between
\citet{pointecouteauetal05} and the Vikhlinin samples, some of which appear in the \citet{planck11_local} sample as well.

To extent that we can, we take care to ensure that all cluster observables are
consistently defined and measured within the same apertures, and rescale all measurements to the same fiducial flat $\Lambda$CDM
cosmology with $\Omega_m=0.3$ and $h=0.7$.  This last rescaling only affects the Vikhlinin measurements, as they adopted $h=0.72$ as their fiducial
Hubble parameter.  Our corrections follow the standard $\Lx \propto h^{-2}$, $M_{gas}\propto h^{-5/2}$, $\Yx \propto h^{-5/2}$, and $M\propto h^{-1}$ scalings.  

This is the first in a pair of papers that characterize the overall systematic differences
in X-ray observables between three independent X-ray analyses, and the impact
that these differences have on cluster scaling relations.   The second paper in the series \citep{rozoetal12c}
utilizes the results from this work to demonstrate that the differences in the scaling relations from X-ray cluster
samples are driven primarily by the systematic differences in the X-ray cluster observables: all additional
sources of complication, including fitting methods and modeling of selection function effects, are sub-dominant
relative to the overall systematic differences in the X-ray observables themselves.  
A third paper \citep{rozoetal12d} extends our analysis to
the optically selected \maxBCG\ cluster catalog \citet{koesteretal07a}, focusing in particular on the tension
noted in \citet{planck11_optical} 
between the observed and expected SZ signal of \maxBCG\ clusters.

The layout of the paper is as follows: Section \ref{sec:data} briefly summarizes each data set
used in our comparison.
Section \ref{sec:xray_comparison} presents the results of our investigation, including
some general discussion in section \ref{sec:xray_comparison_summary}.
Section 4 discusses the surprisingly large amount of evolution in the mass offset in the (P11-LS)--M10 mass comparison.
A summary of the highlights of this work
is presented in section \ref{sec:conclusions}.

Unless otherwise noted, all masses and associated radial and angular scales are defined within a radius 
$\Rf$ that encompasses a mean interior density of $500$ times the critical density of the universe
at the cluster redshift, $\rho_c(z) = 3 H^2(z)/8\pi G$.  
A flat $\Lambda$CDM cosmology with present-epoch matter density $\Omega_m=0.3$ is assumed throughout.


\subsection{A Note on Statistical Errors}
\label{sec:errors}

The core of our analysis consists of comparing a derived cluster property $X$ (e.g. $L_X$, $T_X$, etc) as estimated
in two different works $A$ and $B$.  Given two cluster samples $A$ and $B$, for all clusters in both of these samples
we compute the difference in natural logarithm, 
$\Delta \ln X \equiv \ln( X_{\rm A} /  X_{\rm B} )$.   Averaging over all pairs in common to A and B results in 
a mean difference $\langle \Delta \ln X \rangle_{\rm AB} = r \pm \Delta r$, where $r$ is the
mean offset and $\Delta r$ the $1\sigma$ error in the mean.  Errors are computed via bootstrap resampling, ensuring that the quoted uncertainties
reflects the total variance in the data.

In principle, our analysis should employ the statistical errors of the cluster observables quoted for each sample.  However, we argue that the published uncertainties are unable
to account for the variance seen in sample pairs.  For example, 
using the published errors, the null hypothesis that the mass of a galaxy cluster estimated by \citet{planck11_local}
is consistent with that estimated by \citet{vikhlininetal09} has $\chi^2/dof=307/23$. 
Even after correcting all masses by the mean systematic difference between these two data sets, we
find $\chi^2/dof=75/22$.  This demonstrates that the quoted uncertainties do not account for the full level of variation between the samples,
and explains our reliance on bootstrap resampling for estimating 
uncertainties.\footnote{More generally, if $\epsilon$ is the typical fractional error in $X_i$, then the expected error on the mean $\Delta \ln X_i$ 
estimated from $N$ galaxy clusters is $\sqrt{2/N}  \epsilon$.  The median statistical errors for the clusters under consideration 
have $\epsilon \approx 1\% - 3\%$, so the expected error in the mean is $\Delta r \lesssim 0.01$ (assuming $N=20$), roughly a factor
of two smaller than our direct estimates from bootstrap resampling.}


\section{Cluster Samples}
\label{sec:data}

The  galaxy cluster data used in this work consists of low redshift ($z < 0.3$) clusters drawn from the three samples listed in Table~\ref{tab:samples}.  We describe each sample in turn.  Tables containing the galaxy clusters shared by each pair of cluster samples are detailed in Appendix~\ref{app:data}.


\begin{deluxetable*}{lllll}
\tablewidth{0pt}
\tablecaption{Data Sample Characteristics}
\tablehead{Name &  $N_{\rm cl}$ & X-ray Instrument(s) & Publication(s) & Sample Notes }
\startdata
M10 & 238 & \chandra\ ACIS, \rosat\ PSPC & \citet{mantzetal10a} & Joint cosmology and scaling relation analysis \\
P11-LS &  62 & \xmm\ Epic &  \citet{planck11_local} & Scaling relation analysis; joint SZ+X-ray selection \\
V09 & 85 & \chandra\ ACIS, \rosat\ PSPC & \citet{vikhlininetal06,vikhlininetal09b} & Separate scaling relation and cosmology analyses
\enddata
\label{tab:samples}
\end{deluxetable*}


$\underline{\rm V09}$.  The low redshift clusters in the Vikhlinin sample were selected from
a variety of sources drawn from the \rosat\ All-Sky Survey(RASS) --- the Brightest Cluster Sample \citep[BCS]{ebelingetal00}, 
the \rosat-ESO Flux Limited X-ray sample \citep[REFLEX][]{bohringeretal04},
and the Highest X-ray Flux Galaxy Cluster Sample \citep[HIFLUGCS][]{reiprichbohringer02}.  After applying foreground
cuts, the total area probed was $8.14 \, \mbox{sr}$.  X-ray fluxes were remeasured
using pointed \rosat\ PSPC data when available, and a flux cut of 
$f_X > 1.3\times 10^{11} \mbox{ergs s$^{-1}$ cm$^{-2}$}$ was applied.  
The low redshift sample contains 49 clusters with median estimated mass of 
$\approx 4 \times 10^{14}\ \msun$ that reside primarily at
$z \leq 0.1$, with a few high mass systems out to $z \approx 0.2$.  
All systems have been imaged by \chandra, and these data are
used for estimating cluster temperatures.  For systems
where the \chandra\ field of view is smaller than the cluster radius, \rosat\
data is used for estimating both $\Lx$ and $\Mgas$.  Otherwise, all
data comes from \chandra. 
The cluster masses we use in our comparison are those derived using the $\Mf$--$\Yx$
relation.  This relation is calibrated using relaxed galaxy clusters, and relies on hydrostatic
mass estimates derived from the clusters' X-ray gas and temperature
profile \citep{vikhlininetal06}.

$\underline{\rm M10}$.  The Mantz clusters sample is likewise drawn from three wide-area catalogs from the \rosat\ All-Sky Survey:  BCS, REFLEX, and the bright sub-sample of the Massive Cluster Survey \citep[Bright MACS][]{ebelingetal10}.  Each sample covers a different region of the universe:  BCS contains $z<0.3$ clusters in the Northern sky, REFLEX contains $z<0.3$ systems in the southern sky, and Bright MACS covers $0.3<z<0.5$ at declinations
$>-40^\circ$.  Clusters are selected by applying a redshift-dependent flux cut that approximates a constant luminosity cut corresponding to a mass cut $M\gtrsim 4.7 \times 10^{14}\ \msun$. 
The number of galaxy clusters drawn from each of these samples is 78, 126, and 34.
As in V09, \citet{mantzetal10b} rely on \chandra\ and/or
\rosat\ PSPC pointed data for flux and temperature estimates.  In all cases, \citet{mantzetal10b}
quote \rosat\ calibrated X-ray luminosities and $\Mgas$ values.  For the M10 sample, cluster total masses are estimated from $\Mgas$ based on the $M$--$\Mgas$ scaling relation of \citet{allenetal08}.  This scaling relation was calibrated using hydrostatic masses for relaxed, massive galaxy clusters, where
mass was estimated within the aperture $R_{2500}$ enclosing an overdensity $\Delta=2500$ relative to critical.  The scaling relation
$M_{2500}$--$\Mgas$ was then extended to the outer radius $\Rf$ based on numerical simulations.

$\underline{\mbox{P11-LS}}$.  The \citet{planck11_local} sample is a subset of the Planck Early Sunyaev-Zel'dovich cluster 
sample \citep[ESZ][]{planck11_earlysample}.  The ESZ sample is selected from the Planck all-sky
survey using a multi-frequency, matched filter method \citep{melinetal06} that identifies clusters by their thermal Sunyaev-Zel'dovich (SZ) imprint on the cosmic microwave background.  Clusters are selected by applying a signal-to-noise cut $S/N>6$,
and it is also required that the clusters be identified by at least one additional independent SZ-cluster finding algorithm.
The resulting galaxy clusters are cross-correlated against the Meta-Catalog of X-ray Clusters
\citep[MCXC][]{piffarettietal11}, resulting in 158 cluster matches.  These clusters are searched for in the \XMM\ 
science archive, resulting in 88 matches.  After removing clusters contaminated by flares, or excessively asymmetric galaxy
clusters, the final cluster sample comprises 62 galaxy clusters.  The galaxy clusters cover the redshift range $z\lesssim 0.35$,
with an additional 4 galaxy clusters in the redshift range $z\in[0.35,0.5]$.   The 
selection function described in \citet{planck11_earlysample} has a median mass 
that increases with redshift.
All X-ray quantities --- $\Lx$, $\Mgas$, $\Tx$, and $\Yx$ ---
are estimated from the archival \XMM\ data.   Total masses are estimated from $\Yx$ by relying on the $\Mf$--$\Yx$ scaling relation of \citet{arnaudetal10},
which is calibrated using hydrostatic mass estimates of relaxed galaxy clusters.

Neither M10 nor P11-LS classify the systems in their samples as relaxed/non-relaxed.  However, 
P11-LS does label cool-core clusters, which is generally considered a good proxy for relaxed systems.    
For comparisons involving the V09
sample, we employ the relaxed/unrelaxed characterization of that work to investigate trends with dynamical state.  
For the M10-P11-LS comparison, we employ the cool core characterization of P11-LS.


\section{Comparison of Derived Cluster Properties}
\label{sec:xray_comparison}

In this section, we compare properties for the clusters found in common among pairs of the samples introduced above. 
Since all observables are reported within an aperture $\Rf$ that is estimated by each individual group, it is imperative that we
first estimate any systematic differences in mass calibration, so that we may properly correct all quantities to a common
aperture before we perform the pairwise comparison.  After characterizing differences in mass calibration, we turn
to the remaining cluster observables: $\Lx$, $\Mgas$, $\Tx$, and $\Yx$.

Readers wishing to avoid this level of detail can examine Table~\ref{tab:ratios} and 
skip to section \ref{sec:xray_comparison_summary}, where we summarize our findings.


\begin{figure*}
\begin{center}
\scalebox{0.9}{\plotone{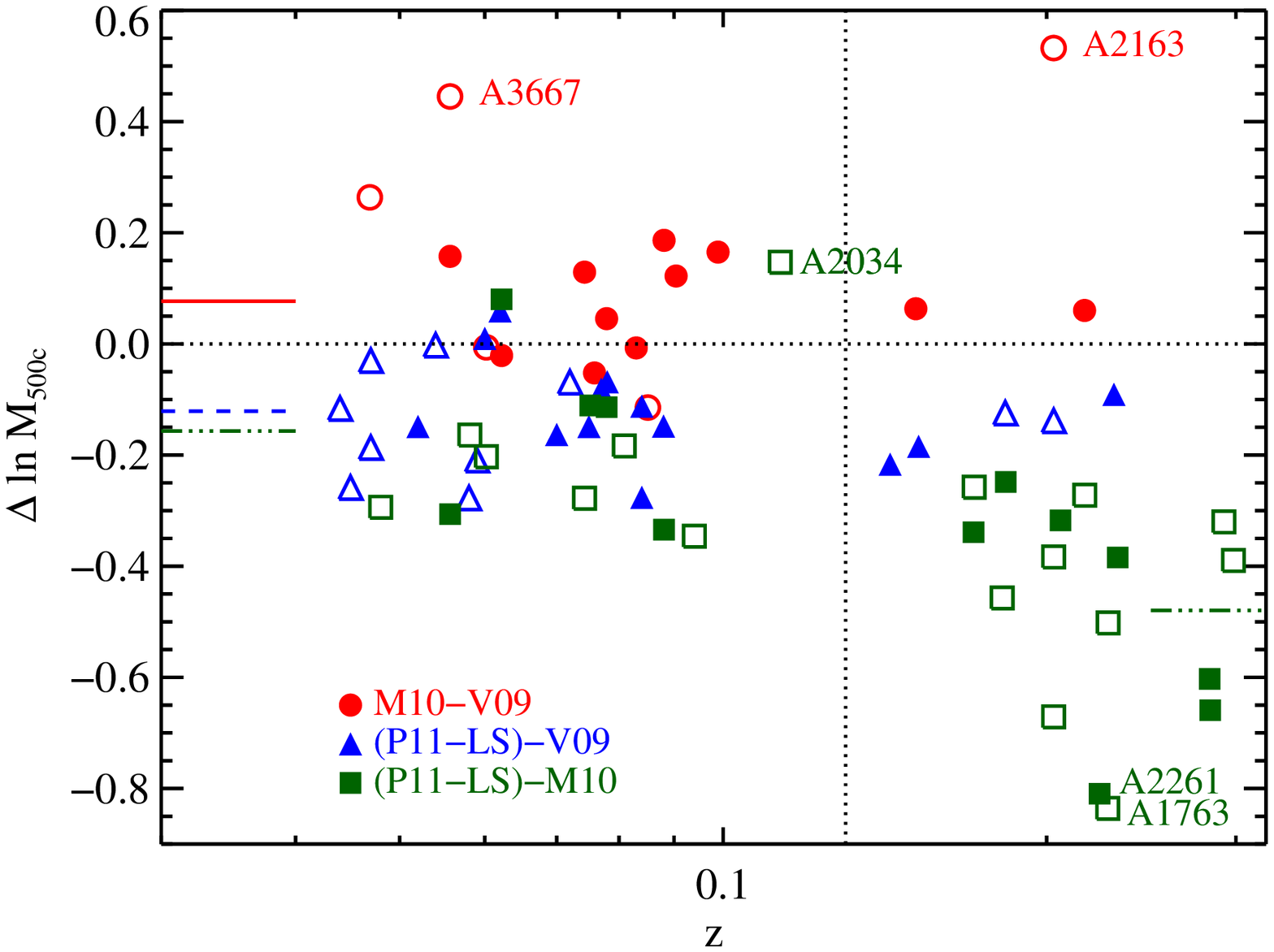}}
\end{center}
\vspace{-0.6truecm}
\caption{Differences in $\Mf$ between clusters shared by the \citet[][V09]{vikhlininetal09}, 
\citet[][M10]{mantzetal10b}, and \citet[][P11-LS]{planck11_local} samples.  
Filled circles are relaxed/cool-core clusters, while open circles are non-relaxed or non cool-core systems.  The small horizontal lines 
along the $y$ axis mark the average mass offset for each of the three cross-comparisons: M10--V09 (solid red), 
(P11-LS)--V09 (dashed blue), (P11-LS)--M10 (dot-dashed green).  
For the (P11-LS)--M10 comparison, we show averages computed separately for low and high redshift systems, split at $z = 0.13$ (vertical dotted line).  The low and high redshift averages are displaced on the left and right axes respectively.  
All averages are computed using only relaxed/cool-core systems.  Clusters labeled are discussed in \S~\ref{sec:specific}.  
}
\label{fig:mass_comparison}
\end{figure*}


\subsection{Mass Comparison}
\label{sec:mass}

We begin by comparing the mass estimates for the samples of galaxy clusters.   As discussed above, each group calibrates 
mass-observable relations against hydrostatic mass estimates of relaxed systems, then uses these calibrated observables to infer masses for the bulk of the 
objects in their samples.  It is important to keep in mind that the list
of galaxy clusters with direct hydrostatic mass estimates is relatively small, and some calibration clusters are yet published,
so directly comparing the actual hydrostatic mass estimates of individual galaxy clusters is difficult.
Fortunately, comparing the masses derived from observational proxies is sufficient for our purposes.
Indeed,
consider two groups, A and B, that calibrate the mass observables relations, 
$M_{\rm A}$--$X_{\rm A}$ and $M_{\rm B}$--$X_{\rm B}$, using their hydrostatic mass estimates of relaxed
galaxy clusters.  The observables $X_{\rm A}$ and $X_{\rm B}$ need not be the same.  
By definition, the mass proxy $M(X)$ for each group is 
\bea
\ln M_{\rm A}  & \equiv  & \avg{\ln M|X_{\rm A}}  , \\
\ln M_{\rm B}  & \equiv  & \avg{\ln M|X_{\rm B}}  ,
\eea
where the average is computed over the calibration sets of {\it relaxed} galaxy clusters by each group. 
It trivially follows that the average mass offset $\avg{\ \ln M_{\rm A}(X_A) - \ln M_{\rm B}(X_B) \ }$ 
between samples A and B  
is an unbiased estimator of the hydrostatic mass offset, {\it so long as one only averages over relaxed galaxy clusters.}
This restriction is there because the calibration of $M(X)$ is done using only relaxed galaxy clusters.
Consequently, we interpret the mass offsets observed here as hydrostatic mass differences, even though the mass
estimates themselves come from mass--observable relations. 
In Paper II, we will make use of the mass offsets identified here when examining differences in published scaling relations.

Figure~\ref{fig:mass_comparison} shows mass differences for clusters in common between the M10--V09, 
(P11-LS)--V09, and (P11-LS)--M10 sample pairs.  Mean differences computed using 
only relaxed/cool-core clusters  (filled symbols) are listed in Table~\ref{tab:ratios} and shown by short lines in the figure.   
The (P11-LS)--V09 and M10--V09 mass comparisons show modest offsets of 
$\avg{\Delta \ln M}=-0.12\pm 0.02$ and $\avg{\Delta \ln M}=0.08\pm 0.02$
respectively.  The (P11-LS)--M10 mass offset using cool-core systems only is 
$\avg{\Delta \ln M} = -0.35\pm 0.07$, but the individual cluster values are sensitive to redshift, with the largest 
discrepancies above $z=0.13$.  Splitting the sample at $z=0.13$, we find mean offsets 
$\avg{\Delta \ln M}=-0.16\pm 0.07$ and $\avg{\Delta \ln M} = -0.48 \pm 0.07$ 
below and above this redshift, which we hereafter refer to as low and high redshifts, respectively.  
These values differ at the $3.2\sigma$ level.

While it would be useful to determine whether there is relative evolution in the other pairings ((P11-LS)--V09 and M10--V09), the sample overlaps are too small to perform a conclusive test.  

In Paper II, we note that the $\Lx$--$\Mtot$ relation of \citet{prattetal09} relies on the hydrostatic mass estimates of \citet{pointecouteauetal05}.  The latter work has 5 clusters in common with V09.  For this common set, we find a mean mass difference $\avg{\Delta \ln M} = -0.18 \pm 0.05$ (P05--V09), consistent with the $-0.12\pm0.02$ mean for the (P11-LS)--V09 sample pair.


\subsubsection{Non-relaxed or No Cool-Core Clusters}

Above, we restricted ourselves to relaxed/cool-core clusters.  This ensures that the observed mass
offsets are unbiased estimates of the hydrostatic mass offsets in the calibration samples.
Nevertheless,
it is interesting to look at the corresponding offsets for non-relaxed systems, as shown in Figure \ref{fig:mass_comparison}
(open symbols).

The mean mass differences for non-relaxed clusters do not differ substantially from those of relaxed systems.  
For the M10--V09 comparison,we find $\avg{\Delta \ln M} =  0.22\pm 0.11$, compared to $0.08\pm 0.02$ for relaxed systems. 
For the (P11-LS)--V09 comparison, we find $-0.14\pm 0.03$ for non-relaxed versus $-0.12\pm 0.02$ for relaxed.  
For (P11-LS)--M10, the non cool-core value for the full sample is $-0.37\pm 0.05$ compared to $-0.35\pm 0.07$ for cool-core
systems.     The (P11-LS)--M10 mass offset does not include cluster A2034 since it is compromised by a poor redshift estimate
in P11-LS (see section \ref{sec:lx}).


\subsubsection{Specific Clusters}\label{sec:specific}

In the M10--V09 comparison, Abell 3667 
Abell 2163 stand out as having large offsets, $\Delta \ln M \sim  0.5$.  Both of these clusters are well-known merging systems.  The X-ray morphology of A 3667 displays a cold front indicative of a recent merger \citep{vik01}, and the cluster displays diametrically opposed radio relics driven by outgoing post-merger shocks \citep{rottgering97}.   Analyzing roughly 500 spectroscopically-confirmed cluster members, 
\citet{owers09} identify multiple galaxy concentrations, the dominant pair of which align with the merger axis seen at X-ray and radio wavelengths.   

A 2163 is a complex system that involves an East-West merger.  There is also an additional northern 
component lying at nearly the same redshift as the main body, but thought to be physically 
distinct \citep{maurogordato08}.  A 2163 also has a cold front and radio halo \citep{bourdin11}, and 
a multi-modal projected mass map derived from gravitational lensing \citep{okabe11}.

The clusters A 1763 and A 2261 exhibit the largest discrepancy in the (P11-LS)--M10 comparison.   
A 1763 is the dominant member of a binary supercluster 
\citep{biviano11}.   A 2261 has been recently investigated with strong and weak lensing analysis by the CLASH team \citep{coeetal12}, 
who produce several estimates under different modeling conditions.  Their main result corresponds to $\Mf=13.4\times 10^{15}\ \msun$,
which favors the higher value of $14.4  \times 10^{14} \msun$ given by 
M10.
The shape and orientation of A2261's host halo, along with projection of surrounding large-scale structure, are identified as key sources 
of systematic uncertainty in their lensing analysis.  They suggest a worst-case scenario in
which there is a 2:1 axis ratio along the line of sight, resulting in a lower mass, $\Mf = 9.4\times 10^{14}\ \msun$.  This value is still closer in log 
to the M10 mass than it is to the P11-LS mass of $\Mf = 6.4\times 10^{14}\ \msun$.
We also label the cluster A 2034 because it stands out as a clear outlier in the $\Lx$ and $\Mgas$
comparisons of P11-LS and M10.  This can be traced to an incorrect redshift assignment in P11-LS.


\subsection{$\Lx$ Comparison}
\label{sec:lx}

We follow \citet{planck11_local} in defining $\Lx$ as the X-ray luminosity within a cylindrical aperture
$\Rf$ in the $0.1-2.4\ \keV$ rest frame band.  This definition is also adopted by \citet{mantzetal10b} and \citet{prattetal09},
whereas \citet{vikhlininetal09} defines $\Lx$ in the $0.5-2.0\ \keV$ band and integrates within a fixed, 2~Mpc aperture.   
We convert the Vikhlinin luminosities to the $0.1-2.4\ \keV$ band by multiplying by $1/0.62$ \citep{bohringeretal04},
and then again by a factor of 0.96 to rescale to luminosities within 
$R_{500}$.\footnote{This correction factor was estimated independently by both A. Vikhlinin and A. Mantz (private communication).
It is somewhat higher than the 0.91 factor advocated by \citet{piffarettietal11}, likely reflecting the fact that they integrate their model out to $5\Rf$, larger than the \citet{vikhlininetal09} scale.
}
Because the mass calibration is different for each of the three works, we should in principle correct the luminosities within $\Rf$
to a common aperture.  In practice, however, 
the dependence of $\Lx$ on aperture is sufficiently weak that the corresponding systematic offsets
are completely negligible (see Appendix \ref{app:aperture}).

Figure \ref{fig:lx_comparison} shows our M10--V09, (P11-LS)--V09, and (P11-LS)--M10 $\Lx$ comparisons, 
as labelled.  There is one gross outlier in this comparison: Abell 2034, which we do not include when estimating
the offset between the various works.
After outlier removal, 
we find $\avg{\Delta \ln \Lx} = -0.01\pm 0.02$ for the (P11-LS)--V09 comparison,  $0.12\pm 0.02$
for M10--V09, and $-0.11\pm 0.02$ for (P11-LS)--M10.
A similar cluster-by-cluster comparison of $\Lx$ for high-redshift systems between M10 and V09 galaxy clusters
shows no evidence of offset in $\Lx$ (Mantz, private communication), suggesting that the V09--M10 difference
observed here may be due to the treatment of \rosat\ data.  Table~\ref{tab:ratios} summarizes the mean offset values. 

A2034 is an outlier because different redshifts are employed by 
P11-LS ($z=0.151$) and M10 ($z=0.113$).  The difference in redshift is $\Delta z/z \approx 0.3$, which induces
an error $\Delta \ln \Lx \approx 0.6$.  Examination of spectroscopic galaxies in the SDSS
makes it apparent that the lower of the two redshifts is the better value;  we find $z=0.1137\pm 0.0008$ 
using 31 galaxies with spectroscopic redshifts within 5 arcmin of the central galaxy. 
Figure \ref{fig:lx_comparison} shows in purple the luminosity offset for A2034
after correcting for the difference in luminosity distances to this object (ignoring the corresponding k-correction).  


\begin{figure}
\begin{center}
\scalebox{1.2}{\plotone{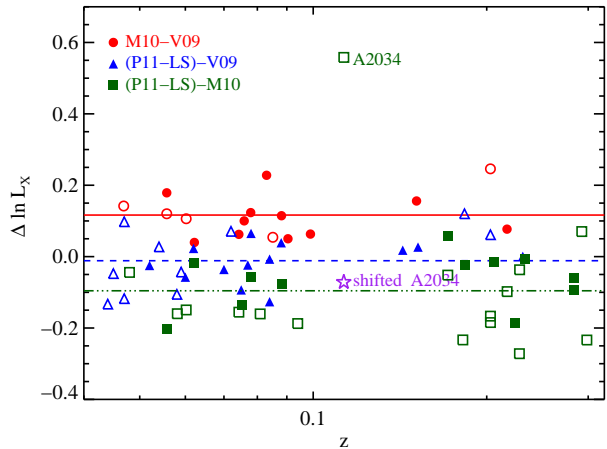}}
\caption{As Figure \ref{fig:mass_comparison}, but now comparing X-ray luminosity, $\Lx$, for clusters in pairs of samples.   
The shifted A2034 point shows the $\Lx$ difference after correcting the P11-LS value to the M10 redshift.
}
\label{fig:lx_comparison}
\end{center}
\end{figure}



\subsection{$\Mgas$ Comparison}
\label{sec:mgas}

\begin{figure}
\begin{center}
\scalebox{1.2}{\plotone{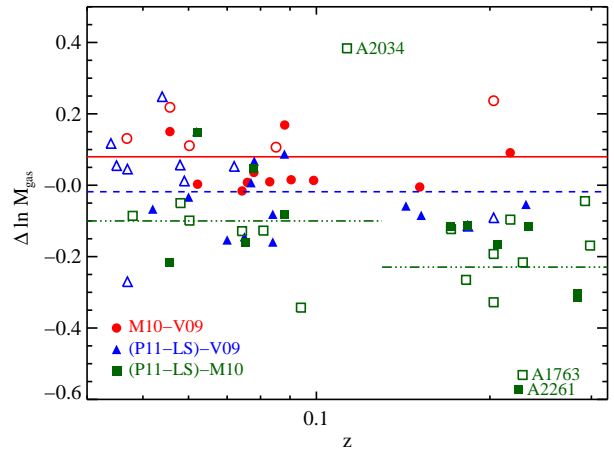}}
\scalebox{1.2}{\plotone{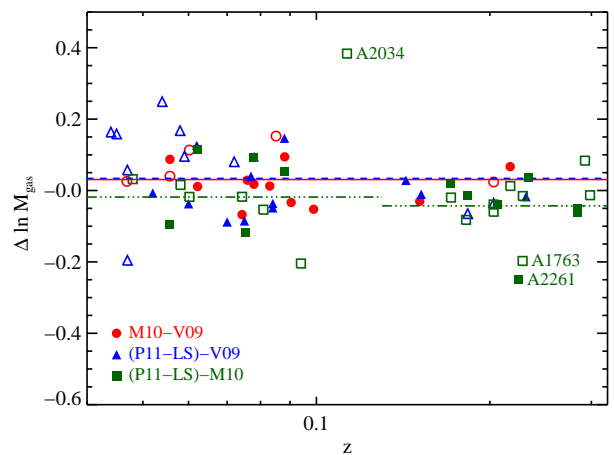}}
\caption{As Figure \ref{fig:mass_comparison}, but now comparing $\Mgas$.  The top panel
shows the raw $\Mgas$ offsets, while the bottom panel accounts for
the difference in aperture $\Rf$ coming from the systematic mass offsets.
Cluster Abell 2034 is excluded from the estimate of the mean $\Mgas$ offset as per the discussion in section \ref{sec:lx}.
Clusters Abell 2261 and Abell 1763 exhibit the largest $\Mgas$ offsets, but are not statistical
outliers in $\Mgas$ after applying the aperture corrections.
}
\label{fig:mgas_comparison}
\end{center}
\end{figure}

The top panel in Figure \ref{fig:mgas_comparison} shows differences in published values of $\Mgas$, the gas mass within a sphere of radius $R_{500}$,  between the various works.   These values are not yet corrected to 
a common aperture.\footnote{\citet{vikhlininetal09}
does not report $\Mgas$ directly, but rather $\Mf$ as estimated from $\Mgas$.  We use the reported $\fgas$--$M$ relation of that work to recover the original $\Mgas$ values.}  Abell 2034 is again an outlier, so we do not include it in any of our calculations.  
Abell 2261 ($\Delta \ln \Mgas = -0.41$) and Abell 1763 ($\Delta \ln \Mgas = -0.37$) 
appear to have unusually low $\Delta \ln \Mgas$ values in the (P11-LS)--M10 comparison.  However, 
using the absolute median deviation to estimate a Gaussian width in the presence of 
outliers\footnote{For a Gaussian distribution, the median absolute deviation $d$ is related 
to the standard deviation via $\sigma=1.4826d$.}, we find that these clusters are $2.7\sigma$ and $2.4\sigma$ away from zero.
Consequently, we do not consider them outliers.


After removing Abell 2034, the M10--V09 and (P11-LS)--V09 mean offsets are $\avg{\Delta \ln \Mgas} =0.08\pm 0.02$ and $-0.02\pm0.02$ respectively. The (P11-LS)--M10 comparison reveals there is redshift evolution in the mean offset between
the two works that mirrors the total mass behavior of section \ref{sec:mass}, with $\avg{\Delta \ln \Mgas}=-0.10\pm0.04$ and $-0.23\pm0.04$
at low and high redshift respectively.    As we now show, these differences primarily reflect the differences in apertures
induced by the mass differences between the three works.

As detailed in Appendix \ref{app:aperture}, a bias $b_M$ in total mass induces a bias in the gas mass within $\Rf$ that 
roughly scales as $b_M^{0.4}$.  To estimate $\Mgas$ using a common aperture for each cluster, we subtract the 
geometrically induced offset $\Delta \ln \Mgas = 0.4\Delta \ln M$ 
from the observed values.  This correction is applied on a cluster-by-cluster basis.

The bottom panel in Figure \ref{fig:mgas_comparison} shows the $\Mgas$ comparison after applying this systematic aperture correction.   
The corresponding $\avg{\Delta \ln \Mgas}$ offsets, listed in Table~\ref{tab:ratios}, are $0.03\pm 0.02$ (M10--V09), 
$0.03\pm 0.02$ ((P11-LS)--V09), 
$-0.02\pm 0.03$ ((P11-LS)--M10, low z), and $-0.04\pm 0.02$ ((P11-LS)--M10, high z).  
There is no evidence of systematic offsets after the aperture correction is applied, demonstrating that
$\Mgas$ is a quantity that is robustly measured by the groups.  
This finding will play a key role in our discussion in section \ref{sec:m_evol}.


\subsection{$\Tx$ Comparison}
\label{sec:tx}


\begin{figure}
\begin{center}
\scalebox{1.2}{\plotone{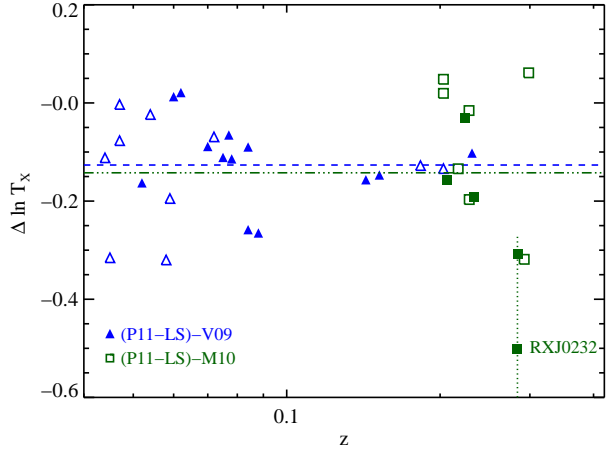}}
\caption{As Figure \ref{fig:mass_comparison}, but now comparing $\Tx$.  There is no M10--V09 comparison because there are only two clusters in the overlap sample with temperature estimates from \citet{mantzetal10b}.  The (P11-LS)--M10 comparison is limited to the subset of 12 clusters that have temperature estimates from \citet{mantzetal10b}.  
}
\label{fig:tx_comparison}
\end{center}
\end{figure}


We define $\Tx$ as the spectroscopic temperature estimated within
a cylindrical annulus of radius $R\in[0.15, 1.0] R_{500}$.  This matches the Mantz and Vikhlinin definition,
but differs from that of Planck, who use an annulus of radius $R\in[0.15,0.75]\Rf$.   
We must correct for this difference in aperture before we perform our comparison.  To do so, we use
temperature estimates within both radial scales for the REXCESS sample of \citet{prattetal09}.  Their data imply a modest correction of the form
\be
\frac{\Tx( [0.15,1]\Rf )}{\Tx ( [0.15,0.75]\Rf )} = 0.95 \left( \frac{\Tx( [0.15,0.75]\Rf )}{5.0\ \keV} \right)^{0.016}.
\ee
We use this relation to correct all of their $\Tx$ values to a $[0.15,1]\Rf$ aperture.
As we were completing this work, \citet{rozoetal12a} argued that the
temperature corrections calibrated from the \citet{prattetal09} data are not correct, and that one incurs a greater error
by applying this correction than by simply ignoring the difference in the definition of $T_X$.  Using deeper \xmm\ data,
Pratt (private communication) estimates the temperature ratio at 0.97 rather than 0.95.   We will ignore this small difference
in our discussion, and simply note that the temperature correction we apply is likely too large by $\approx 2\%-3\%$.

There are other slight differences in the temperature definitions between the three groups that we do not attempt to account for.  The spectral energy range used to derive X-ray temperatures are subtly different:  [0.6--10], [0.8--7.0], and [0.3--10]~keV  for V09, M10, and P11-LS, respectively.  Simulations \citep{mathiesen01} and observations \citep{cavagnolo08} suggest that derived temperatures depend on bandpass, with harder spectral ranges yielding generally higher temperatures.  However, we expect these effects should be small for the subtle bandpass differences between V09, M10 and P11-LS. 
In addition, in principle, we should also correct all temperatures to ensure a consistent choice of $\Rf$.  In practice, however,
the dependence of $\Tx$ on $\Rf$ is weak enough that it may be safely neglected (see Appendix \ref{app:aperture}).  

Figure \ref{fig:tx_comparison} shows the (P11-LS)--V09 and (P11-LS)--M10 temperature comparisons.  
The mean offsets of $-0.13 \pm 0.02$ and $-0.14\pm 0.05$ respectively, are listed in Table~\ref{tab:ratios}.  
In the (P11-LS)--M10 comparison we limit ourselves to the subsample of 12 galaxy clusters that have independently estimated temperatures.
We do not show a M10--V09 comparison because only 6 of the
16 clusters in the overlap sample have independent temperature measurements, and 4 of those are from \citet{horner01},
which are neither core nor clump excised.  The cluster RX-J0232.2, observed by P11-LS and M10, 
exhibits the largest discrepancy, but the large error in the M10 temperature of $\Tx = 10.06 \pm 2.31\ \keV$ means 
that this difference is not significant.

Differences similar to the $13\%$ offset we find between \chandra\ and \XMM\ temperatures 
have been noticed before, and are typically attributed to instrument calibration uncertainties \citep[e.g.][]{vikhlininetal05, prattetal09}.  For example, 
\citet{nevalainenetal10} find instrumental calibration can lead to systematic offsets 
$\Delta \ln T= -0.08$ (ACIS/MOS1) or $\Delta \ln T=-0.14$ (ACIS/MOS1 and ACIS/MOS2).
Interpretation of our results as temperature calibration offsets, however, is not trivial, as the  \chandra\  calibration 
version employed by \citet{vikhlininetal09} is different from that of \citet{nevalainenetal10}.   We note also 
that \citet{nevalainenetal10} employ a spectral bandpass of $[0.5,7]\keV$, slightly different from the choices 
of the other three groups. 

In addition, it is also clear that differences in substructure masking plays a role in at least a subset of the galaxy clusters under consideration.
In particular, the two largest outliers in the (P11-LS)--V09 comparison are Abell 3376 and Abell 2256.
Both of these systems are known to have large, cool substructures, which the final temperature estimates are sensitive to.  
A comparison of the masks used to estimate $\Tx$ in \citet{planck11_local}
and \citet{vikhlininetal09} reveals that there the two masks are different
(A. Vikhlinin and H. Bourdin, private communication).  A reanalysis of the X-ray data of these two clusters using
the same masks results in significantly better agreement between the two works 
(A. Vikhlinin and H. Bourdin, private communication).


\subsection{$\Yx$ Comparison}
\label{sec:yx}

\begin{figure}
\begin{center}
\scalebox{1.2}{\plotone{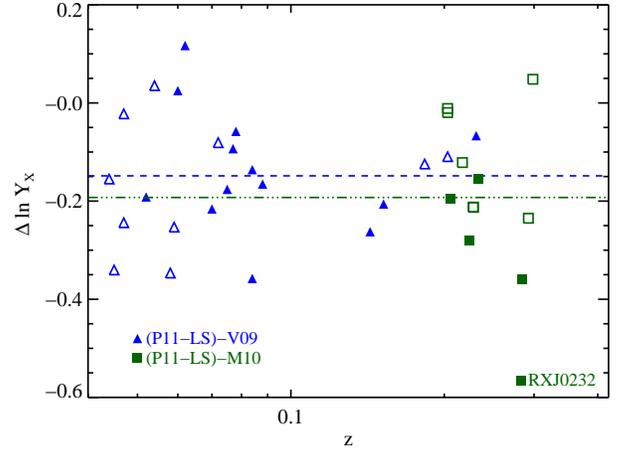}}
\caption{As Figure \ref{fig:mass_comparison}, but now comparing $\Yx$.  The gas mass has been corrected for the difference in the aperture $\Rf$ based on the results from Figure \ref{fig:mass_comparison}.   
}
\label{fig:yx_comparison}
\end{center}
\end{figure}

The X-ray gas thermal energy, $\Yx$, is defined as the product $\Yx= \Mgas \Tx$.  
Figure \ref{fig:yx_comparison} compares the $\Yx$ estimates of galaxy clusters between the various works.  
We have again corrected the \citet{planck11_local} temperatures to match the definition of \citet{vikhlininetal09},
and we have corrected the $\Mgas$ values to a common metric aperture as per section \ref{sec:mgas}.
We do not include a M10--V09 comparison, since there are only two clusters with independent core-excised temperature measurements.

We find $\avg{\Delta \ln \Yx} = -0.15\pm 0.03$  for the (P11-LS)--V09 difference and $-0.19 \pm 0.05$ for (P11-LS)--M10.    The (P11-LS)--M10 offset
is essentially in perfect agreement with the expectation that $\avg{\Delta \ln \Yx} = \avg{\Delta \ln \Mgas} + \avg{\Delta \ln T_X}$,
whereas for the (P11-LS)--V09 comparison, we have $ \avg{\Delta \ln \Mgas} + \avg{\Delta \ln T_X} = -0.10\pm 0.03$, which suggests
the $\Mgas$ and $T_X$ offsets of P11-LS and M10 are correlated.   The most discrepant system is again RX-J0232.2, but the
error in the X-ray temperature in M10 for this system is large.


\subsection{Summary}
\label{sec:xray_comparison_summary}


\begin{deluxetable*}{lllll}
\tablewidth{0pt}
\tablecaption{Mean Log Differences in X-ray Properties for Sample Pairs}
\tablehead{Property & M10 -- V09 & P11-LS -- V09 & P11-LS -- M10  & P11-LS -- M10  \\
& & & Low z ($z\leq 0.13$) & High z ($z>0.13$)}
\startdata
$\Lx$\footnote{Offset computed after outlier removal.} & $0.12\pm 0.02$ & $-0.01\pm 0.02$ & $-0.12\pm 0.02$ & $-0.10\pm 0.03$ \\
$\Mgas$$^a$\footnote{Offset computed after correction to a common aperture.} & $0.03\pm 0.02$ &  $0.03\pm 0.02$ & $-0.02\pm 0.03$ & $-0.04\pm 0.02$  \\
$\Tx$ & --- & $-0.13\pm 0.02$ & --- &  $-0.14\pm 0.05$ \\
$\Yx$$^{ab}$ & --- & $-0.15\pm 0.03$ & --- & $ -0.19 \pm 0.05$ \\
$\Mtot$\footnote{Relaxed/cool core only.} & $0.08 \pm 0.02$ &  $-0.12\pm 0.02$ & $-0.16\pm 0.07$ & $-0.48\pm 0.07$ \\
$\Mtot$\footnote{Non-relaxed/no cool core only.}  & $0.22 \pm 0.11$ &  $-0.14\pm0.03$  & $-0.25\pm 0.03$  & $-0.45\pm0.06$ 
\enddata
\label{tab:ratios}
\end{deluxetable*}


Table~\ref{tab:ratios} summarizes the mean logarithmic offsets for each of the cluster properties considered in this work, with errors in the mean values derived from bootstrap resampling of the clusters in common between each pair of samples.  

We find significant systematic offsets in total mass estimates between the various groups, being as large as $20\%$ at low redshift, 
and growing to $\approx 45\%$ at $z\approx 0.2$ for the (P11-LS)--M10 comparison.   We re-emphasize 
that this comparison necessarily rests on mass proxies (rescaled observables) rather than direct hydrostatic mass estimates.  However, 
we have argued that this mass offset should offer an unbiased estimate of the hydrostatic masses applied by the different groups, since each uses hydrostatic masses to calibrate their relevant observable--mass scaling relation.  
Future work comparing explicit hydrostatic masses across different groups is highly desirable.

After applying an aperture correction based on total mass estimates, we find that all groups are in good agreement when it comes to 
estimating $\Mgas$.  The level of systematic differences in the mean are only a few percent, consistent with the overall systematic 
uncertainty between \chandra\ and \xmm\ cross-calibration in the soft X-ray band, considered to be $\approx 10\%$ in flux or $5\%$ in $\Mgas$ \citep{nevalainenetal10,tsujimotoetal11}.  

The good agreement in $\Mgas$ is not entirely reflected in the $\Lx$ comparison, where we find $12\%$ systematic offsets in the M10--V09 and low redshift (P11-LS)--M10 comparisons.   
These offset are not simply due to instrumental calibration, since otherwise one should  have $\Delta \ln \Mgas = 0.5\Delta \ln L_X$, which is
not satisfied.     Detailed comparative studies, in which the same cluster observations are independently analyzed by different groups, are needed to clarify the origins  of the $\Lx$ differences found here. 

Consistent with previous calibration studies such as \citet{nevalainenetal10}, we find $\sim 13\%$ differences in mean temperature estimates of clusters.
The principal source of these offsets is not clear: it could reflect instrumental calibration, or differences in the data analysis (e.g. substructure masking).
Differences in gas thermal energy, $\Yx$, are consistent with the product of the differences in $\Mgas$ and $\Tx$ for the (P11-LS)--M10 comparison,
but there may be differences in the (P11-LS)--V09 comparison.  This difference 
may be evidence that the $\Mgas$ and $T_X$ systematic offsets are correlated.


\section{Redshift Behavior of (P11-LS)--M10 Estimates}
\label{sec:m_evol}

In the analysis of $\Mtot$ above, the (P11-LS)--M10 values display redshift sensitivity, with larger discrepancy for $z > 0.13$.   
A detailed discussion of possible interpretations of this result is presented in Appendix \ref{app:m_evol}.   Here, we present only a summary
of the discussion, since the details are cumbersome.  
We first verity that the evolution in the mass ratio is significant by fitting a power-law model 
$\Delta \ln M \propto \gamma \ln(1+z)$, which requires $\gamma = -1.90 \pm 0.44$, indicating evolution at high confidence.   
We attempt to ascertain which of the two data sets is driving this evolution.   Specifically, we first assume that all evolution
is due to the M10 data set, and then consider the observational implications of this hypothesis.  We then consider the converse hypothesis --- 
that all evolution is due to systematics in the P11-LS data set --- 
and consider its observational implications.  We are not able to rule out either hypothesis, but we present the basic
results and suggest further work to clarify the source of the discrepancy.  
Of course, the true answer may lie between the two simple extremes posed here.  

Suppose that the observed evolution is due solely to systematics in the M10 data set.  M10 adopted a constant
$\fgas$ model, deriving $\Mtot$ from $\Mgas$ by assuming a fixed gas fraction value, $f_0$.   Since all groups
agree on $\Mgas(R)$ within a common aperture, the only problem that can arise with the M10 masses is if their constant $\fgas$
assumption is incorrect.  Given an arbitrary $\fgas$ model, the bias induced by assuming $\fgas=f_0$ 
is
\be
\Delta \ln M = 1.67\ln \left( \frac{\fgas(M,z)}{f_0} \right)  , 
\ee
where the value of $1.67$ arises from the aperture correction discussed in \S~\ref{sec:mgas}.  
We have verified that if one uses an $\fgas$ model derived from the P11-LS data, one can 
reproduce the mass-offset behavior seen in Figure~\ref{fig:mass_comparison}.  Note in replacing
the $\fgas$ model in this way, the agreement in $M_{500}$ simply reflects the agreement in $M_{gas}$
between the two works.

The redshift and mass dependence of the gas mass fraction is an active topic of investigation.  
Most works to date \citep[][]{vikhlininetal09,sunetal09,prattetal09} argue that the gas
mass fraction $\fgas \propto M^{\alpha_g}$ where $\alpha_g\approx 0.1-0.2$.  In addition,
a recent study by \citet{linetal11} argue not only for mass scaling ($\alpha_g\approx 0.13$),
but also redshift scaling.  On the other hand, \citet{allenetal08} --- on which the work of
\citet{mantzetal10b} is based --- find no scaling of $\fgas$ with mass and/or redshift within the range probed by their data.
In short, whether or not the M10 masses are correct depends on the slope of the $\fgas$--$M$ relation, and 
there are plausible models for the $\fgas$--$M$ relation that can induce bias in the M10 masses.  
Whether such a bias is real or not, however, requires a better understanding of the gas mass fraction $\fgas(M,z)$.

We now consider the alternate proposition, namely that the P11-LS masses are biased.  The good agreement
in $\Mgas$ between M10 and P11-LS suggest that the total mass bias should be sourced primarily by cluster temperature. 
Thus, if the M10 masses are unbiased, the evolving bias in the (P11-LS)--M10 comparison implies an evolving temperature
bias in the P11-LS temperatures.  We search for such an evolution by considering 
an effective ``SZ temperature'', $T_{SZ}=\Ysz/\Mgas$, and examine the ratio $T_{SZ}/\Tx \equiv \Ysz/\Yx$ in the P11-LS data 
using the published $\Ysz$ values derived from Planck satellite observations.  
We find that the $\Ysz/\Yx$ ratio of P11-LS exhibits redshift dependence at the $2.7\sigma$ level, suggestive but not definitive.  
We further demonstrate that such an evolution --- if real --- is not sourced by selection effects (at least in the form of $S/N$ cuts), 
but would have to be endemic to the cluster population.

We also note in Appendix \ref{app:m_evol} that 
the mean $\ln (\Ysz/\Yx)$ value for $z\leq 0.13$ clusters is slightly negative ($-0.06 \pm 0.04$) while the value 
for  $z \leq 0.13$ clusters is slightly positive ($0.08 \pm 0.04$).   
The redshift transition lies in the range  that divides the A and B subsets of clusters in 
\citet{planck11_local}.   These samples are characterized as having $\Rf$ values that subtend angular scales greater than or less than $12^\prime$, respectively, 
where  $12^\prime$ is the maximum angular scale within which the X-ray background can be determined in a single \xmm\ field of view.  
Thus, it is possible that the redshift evolution we observe simply reflects systematic differences inherent to the A and B cluster populations.
The most obvious possibility is background subtraction, as the background treatment
for A and B clusters is different.
That said,  as noted by \citet{planck11_local}, the set of clusters in P11-LS is neither representative nor complete, so we heed
that caution, and leave it to future work to further pursue the origin of the systematic difference between A and B clusters in P11-LS.

We note, however, that the ratio $\Ysz/\Yx = 1.08 \pm 0.04$ seen above $z = 0.13$ is unusually high.  
This result in in strong ($6.3\sigma$) tension with that of \citet{rozoetal12a} for $z\approx 0.1$ galaxy cluster using
{\it Planck} and \chandra\ data.  What is most 
surprising about this result, however, is that it appears to be in conflict with X-ray expectations not only at the quantitative
but even at the qualitative level.  Specifically,
galaxy clusters are known to have falling temperature profiles \citep[e.g][]{vikhlininetal06,arnaudetal10}.
Compared to SZ measurements, X-ray temperature weigh the inner, hotter (remember these are core-excised temperature) 
regions of a galaxy cluster more heavily than the outskirts.  This in turn
leads one to a generic prediction that $\Ysz/Y_X\leq 1$, a prediction that is only strengthened by the presence of unresolved
gas clumping.   It seems difficult to reconcile this X-ray expectation to the $\Ysz/\Yx = 1.08 \pm 0.04$ seen above $z = 0.13$.


\section{Summary and Conclusions}
\label{sec:conclusions}

We compare recent estimates of cluster properties within $\Rf$ published by three independent groups based on \chandra\ and \xmm\ data.  Using clusters in common to pairs of studies, we form ratios of the estimated total masses, gas masses, temperatures and gas thermal energies for individual clusters.   Mean values of these ratios are used to gauge the level of systematic error in X-ray analysis.   Our findings are summarized as follows.
\\

$\bullet$  Total masses derived from scaled observables differ at the 10\% to nearly 50\% depending on the two works being compared,
and the redshift range sampled.  These differences exist even for relaxed and cool-core clusters.\\

In the (P11-LS)--M10 comparison of 16 systems at $z > 0.13$, total masses deviate by $48\% \pm 7 \%$ in the mean, with A2261 and A1763 having individual mass estimates that differ by more than a factor of two between the two groups.  
Similar differences have been noted before \citep[see][Appendix B]{rykoffetal12}, but this work is the first to attempt to quantify these systematic offsets between statistically relevant clusters samples drawn from the literature.  As we demonstrate in paper II, these differences are the primary source of tension between the cluster scaling relations derived by the three groups we consider.
\\

$\bullet$ $\Mgas$ within a common scaled aperture is a quantity that is robustly estimated.\\

If the behavior of the gas mass fraction, $\fgas(M,z)$, were known, this result suggests $\Mgas$ would be the most observationally robust X-ray mass proxy.  
There remains debate as to the gas fraction behavior.   While most studies in the literature find that $\fgas$ scales with mass as $M^{0.1-0.2}$ 
\citep{vikhlininetal09,prattetal09,sunetal09,linetal11}, there is also evidence that the scaling saturates to a constant value for the most massive galaxy clusters \citep{allenetal08}.  Full convergence on a unique solution has not yet been reached, and current differences may simply reflect overall
differences in the mass range sampled by the various galaxy cluster samples.
\\

$\bullet$   There are $5\%-15\%$ systematic differences in $\Lx$, $\Tx$ and $\Yx$ between the various groups. \\

The deviations in $\Yx$ are consistent with the product of the $\Mgas$ and $\Tx$ deviations in the (P11-LS)--M10 comparison, 
but less so for the (P11-LS)--V09 comparison, suggesting correlated offsets.
The primary source of these discrepancies is unclear: neither instrumental calibration \citep{nevalainenetal10,tsujimotoetal11} 
nor methodological differences can currently be ruled out, and both appear to contribute at some level.  
Resolving this problem requires a concerted program to explicitly test data reduction pipelines, using both real and simulated data, 
in a controlled environment.
\\

$\bullet$ There is apparent redshift evolution in the relative (P11-LS)--M10 mass calibration, with strong tension between the two works
at $z>0.13$.\\  

We show that modifying the \citet{mantzetal10b} masses to incorporate a mild trend in $\fgas$ with mass can 
account for the differences seen, but an alternative scenario involving redshift evolution in P11-LS mass estimates 
is also possible.  In the latter case, the $\Ysz/\Yx$ ratio should also evolve with redshift, and such evolution is in fact 
seen at modest  ($2.7\sigma$) significance.   The average value of $\Ysz/\Yx = 1.08\pm 0.04$ for $z > 0.13$ galaxy clusters is 
high compared to recent \chandra\ measurements of $0.82 \pm 0.02$ \citep{rozoetal12a}, and difficult to reconcile 
with the well-established falling temperature
profiles of galaxy clusters.

X-ray-based cluster methods have a long and storied history as cosmological probes, and the cosmological constraints from cluster samples today are competitive with, and complementary to, the best large-scale structure methods available.  Even if weak lensing methods of mass calibration eventually become more robust, thanks to low-scatter mass proxies such as $\Mgas$ and $\Yx$, X-ray observations are guaranteed to remain invaluable in helping improve cosmological constraints in future cluster samples.   Realizing the full promise of the cluster population, however, requires that the systematic differences highlighted in this work be fully resolved.  

\acknowledgements The authors would like to thank Adam Mantz, Alexey Vikhlinin, Gabriel Pratt, Monique Arnaud, 
and Steven Allen for useful criticisms 
on earlier drafts of this work.  The authors would also like to thank the organizers of the Monsters Inc. workshop at KITP, 
supported in part by the National Science Foundation under Grant No. PHY05-51164, 
where this collaboration was started.  ER gratefully acknowledges the hospitality of the AstroParticle and Cosmology 
laboratory (APC) at the Universit\'e Paris Diderot, where part of this work took place. 
ER is funded by NASA through the Einstein Fellowship Program, grant PF9-00068.  
AEE acknowledges support from NSF AST-0708150 and NASA NNX07AN58G.  
A portion of the research described in this paper was carried
out at the Jet Propulsion Laboratory, California Institute of Technology, under a
contract with the National Aeronautics and Space Administration.
This work was supported in part by the U.S. Department of Energy contract to SLAC no. DE-AC02-76SF00515.

\newcommand\AAA[3]{{A\& A} {\bf #1}, #2 (#3)}
\newcommand\PhysRep[3]{{Physics Reports} {\bf #1}, #2 (#3)}
\newcommand\ApJ[3]{ {ApJ} {\bf #1}, #2 (#3) }
\newcommand\PhysRevD[3]{ {Phys. Rev. D} {\bf #1}, #2 (#3) }
\newcommand\PhysRevLet[3]{ {Physics Review Letters} {\bf #1}, #2 (#3) }
\newcommand\MNRAS[3]{{MNRAS} {\bf #1}, #2 (#3)}
\newcommand\PhysLet[3]{{Physics Letters} {\bf B#1}, #2 (#3)}
\newcommand\AJ[3]{ {AJ} {\bf #1}, #2 (#3) }
\newcommand\aph{astro-ph/}
\newcommand\AREVAA[3]{{Ann. Rev. A.\& A.} {\bf #1}, #2 (#3)}

\bibliographystyle{apj}
\bibliography{mybib}

\appendix


\section{A:  Sample Comparison Data}
\label{app:data}

The data employed in our comparisons is presented here.  Table~\ref{tab:mv_sample} lists clusters common to the M10--V09 samples, Table~\ref{tab:pv_sample} the (P11-LS)--V09 samples, and Table~\ref{tab:pm_sample} the (P11-LS)--M10 samples.  As per the discussion in section \ref{sec:errors}, we report only central values, since the quoted uncertainties cannot explain the systematic shifts in most properties.  Data compiled in these tables are raw values extracted from the relevant papers \citep{vikhlininetal06, vikhlininetal09, mantzetal10b,planck11_local}, and before applying any of the corrections described in the text.   

Units are as follows: $\Lx\ (10^{44}\ \mbox{ergs/s})$, $\Mgas\ (10^{14}\ \msun)$,
$\Tx\ (\keV)$, $\Yx\ (10^{14}\ \keV\msun)$, $C^{-1}D_A^2 \Ysz\ (10^{14}\ \keV\msun)$.  We assume
$C=1.416\times 10^{-19}\ \Mpc^2 \keV^{-1} \msun^{-1}$ as per \citet{arnaudetal10}.


\begin{deluxetable*}{lccccccc}
\tablewidth{0pt}
\tablecaption{M10--V09 Cluster Sample}
\tablehead{Name & $z$ & $\Lx^{Vik}$ & $\Lx^{Mantz}$ & $\Mgas^{Vik}$ & $\Mgas^{Mantz}$  & $\Mf^{Vik}$ & $\Mf^{Mantz}$ }
\startdata
A85 & 0.0557 & 2.91 & 5.70 & 0.69 & 0.82 & 5.98 & 7.20 \\
A401 & 0.0743 & 3.90 & 6.80 & 1.15 & 1.16 & 8.63 & 10.10 \\
A478 & 0.0881 & 7.24 & 13.30 & 0.95 & 1.15 & 8.15 & 10.10 \\
A1795 & 0.0622 & 3.52 & 6.00 & 0.61 & 0.63 & 5.46 & 5.50 \\
A2029 & 0.0779 & 5.72 & 10.60 & 1.01 & 1.16 & 8.64 & 9.30 \\
A2142 & 0.0904 & 7.20 & 12.40 & 1.52 & 1.59 & 11.96 & 13.90 \\
A2204 & 0.1511 & 9.35 & 17.90 & 1.15 & 1.18 & 9.40 & 10.30 \\
A2244 & 0.0989 & 2.98 & 5.20 & 0.67 & 0.70 & 5.11 & 6.20 \\
A2597 & 0.0830 & 2.09 & 4.30 & 0.32 & 0.33 & 2.84 & 2.90 \\
A3112 & 0.0759 & 2.43 & 4.40 & 0.46 & 0.47 & 4.20 & 4.10 \\
RX J1504 & 0.2169 & 15.60 & 27.60 & 1.11 & 1.25 & 10.07 & 11.00 \\
\\
A2163 & 0.2030 & 13.70 & 28.70 & 3.36 & 4.40 & 21.98 & 38.50 \\
A2256 & 0.0581 & 2.66 & 4.60 & 0.72 & 0.82 & 7.85 & 7.20 \\
A3266 & 0.0602 & 2.69 & 4.90 & 0.92 & 1.06 & 9.00 & 9.20 \\
A3558 & 0.0469 & 1.96 & 3.70 & 0.63 & 0.73 & 4.78 & 6.40  \\
A3667 & 0.0557 & 3.14 & 5.80 & 1.06 & 1.35 & 7.35 & 11.80
\enddata
\label{tab:mv_sample}
\tablecomments{Units are $10^{44}$ ergs/s for $L_X$, $10^{14}\ \msun$ for mass. Precise definitions specified in the text.  
The list is divided by the blank row into relaxed (upper) and merging (lower) systems.   Values listed are raw data drawn from the literature, before applying any of the corrections mentioned in the main body of the text.}
\end{deluxetable*}



\begin{deluxetable*}{lcccccccccccc}
\tablewidth{0pt}
\tablecaption{(P11-LS)--V09 Cluster Sample}
\tablehead{Name & $z$ & $\Lx^{Pl}$ & $\Lx^{Vik}$ & $\Mgas^{Pl}$ & $\Mgas^{Vik}$ & $\Tx^{Pl}$ & $\Tx^{Vik}$ & 
	$\Yx^{Pl}$ & $\Yx^{Vik}$ & $\Mf^{Pl}$ & $\Mf^{Vik}$ & $C^{-1}D_A^2\Ysz$ }
\startdata
A85 & 0.0557 & 4.65 & 2.91 & 0.66 & 0.69 & 5.78 & 6.45 & 3.81 & 4.44 & 5.30 & 5.98 & 3.32 \\ 
A401 & 0.0743 & 5.82 & 3.90 & 1.02 & 1.15 & 7.26 & 7.72 & 7.41 & 8.52 & 7.65 & 8.63 & 5.86 \\ 
A478 & 0.0881 & 12.33 & 7.24 & 1.06 & 0.95 & 6.43 & 7.96 & 6.82 & 7.74 & 7.23 & 8.15 & 6.50 \\ 
A1413 & 0.143 & 3.39 & 4.11 & 0.53 & 0.81 & 6.59 & 7.30 & 3.49 & 6.93 & 4.90 & 7.57 & 4.87 \\
A1650 & 0.0823 & 3.79 & 2.33 & 0.51 & 0.54 & 5.11 & 5.29 & 2.61 & 2.82 & 4.22 & 4.59 & 3.11 \\ 
A1651 & 0.0853 & 4.23 & 2.93 & 0.56 & 0.64 & 5.23 & 6.41 & 2.93 & 4.23 & 4.51 & 5.78 & 2.54 \\ 
A1795 & 0.0622 & 5.90 & 3.52 & 0.73 & 0.61 & 6.60 & 6.14 & 4.82 & 3.80 & 5.96 & 5.46 & 3.25 \\ 
A2029 & 0.0779 & 10.00 & 5.72 & 1.12 & 1.02 & 7.70 & 8.22 & 8.62 & 8.55 & 8.30 & 8.64 & 5.72 \\ 
A2204 & 0.1511 &15.73 & 9.35 & 1.09 & 1.15 & 7.75 & 8.55 & 8.45 & 10.17 & 8.04 & 9.40 & 7.84 \\ 
A2390 & 0.2329 & 17.20 & 10.49 & 1.54 & 1.51 & 8.89 & 9.40 & 13.69 & 13.84 & 10.35 & 11.02 & 11.72 \\ 
A3112 & 0.0759 & 3.84 & 2.43 & 0.40 & 0.46 & 5.02 & 5.19 & 2.01 & 2.40 & 3.67 & 4.20 & 1.27 \\ 
A3158 & 0.0583 & 2.66 & 1.72 & 0.53 & 0.54 & 5.00 & 4.67 & 2.65 & 2.33 & 4.29 & 4.13 & 2.47 \\ 
Zw1215 & 0.0767 & 2.88 & 1.80 & 0.63 & 0.61 & 6.45 & 6.54 & 4.06 & 4.18 & 5.45 & 5.75 & 3.25 \\ 
A1689\footnote{Though relaxed in appearance, A1689 is known to have several structures
superimposed along the line of sight \citep{lokasetal06}.} & 0.1832 & 13.29 & 7.20 & 1.08 & 1.13 & 8.17 & 8.85 & 8.82 & 9.57 & 8.19 & 9.02 & 9.68 \\
\\
A119 & 0.0445 & 1.52 & 1.06 & 0.45 & 0.39 & 5.40 & 5.72 & 2.43 & 2.69 & 4.12 & 4.50 & 1.91 \\ 
A754 & 0.0542 & 4.68 & 2.78 & 1.04 & 0.79 & 8.93 & 8.73 & 9.29 & 8.19 & 8.69 & 8.47 & 6.07 \\ 
A1644 & 0.0475 & 1.66 & 1.14 & 0.41 & 0.53 & 4.86 & 4.61 & 1.99 & 2.48 & 3.66 & 4.29 & 1.77 \\ 
A2065 & 0.0723 & 3.20 & 1.82 & 0.60 & 0.56 & 5.36 & 5.44 & 3.22 & 3.24 & 4.78 & 4.98 & 2.75 \\ 
A2163 & 0.2030 & 23.86 & 13.70 & 3.17 & 3.36 & 13.40 & 14.72 & 42.48 & 45.99 & 19.68 & 21.98 & 1.48 \\ 
A2256 & 0.0581 & 3.92 & 2.66 & 0.78 & 0.72 & 6.40 & 8.37 & 4.99 & 7.16 & 6.11 & 7.84 & 5.01 \\ 
A3266 & 0.0602 & 4.22 & 2.69 & 0.96 & 0.92 & 7.46 & 8.63 & 7.16 & 9.12 & 7.51 & 9.00 & 6.36 \\ 
A3376 & 0.0455 & 0.92 & 0.59 & 0.28 & 0.26 & 3.39 & 4.37 & 0.95 & 1.33 & 2.39 & 3.01 & 0.71 \\ 
A3558 & 0.0557 & 3.54 & 1.96 & 0.67 & 0.63 & 4.78 & 4.88 & 3.20 & 2.99 & 4.77 & 4.78 & 2.97 \\ 
\enddata
\label{tab:pv_sample}
\tablecomments{Units are $10^{44}$ ergs/s for $L_X$, $10^{14}\ \msun$ for mass, $\keV$ for temperature, and $10^{14}\ \msun\keV$ for
$Y_X$ and $C^{-1}D_A^2\Ysz$.  Precise definitions of all quantities are specified in the text.    
The list is divided by the blank row into relaxed (upper) and merging (lower) systems.}
\end{deluxetable*}



\begin{deluxetable*}{lcccccccccccc}
\tablewidth{0pt}
\tablecaption{(P11-LS)--M10 Cluster Sample}
\tablehead{Name & $z$ & $\Lx^{Pl}$ & $\Lx^{Mantz}$ & $\Mgas^{Pl}$ & $\Mgas^{Mantz}$ & $\Tx^{Pl}$ & $\Tx^{Mantz}$ & 
	$\Yx^{Pl}$ & $\Yx^{Mantz}$ & $\Mf^{Pl}$ & $\Mf^{Mantz}$ & $C^{-1}D_A^2\Ysz$ }
\startdata
A85 & 0.0557 & 4.65 & 5.70 & 0.66 & 0.82 & 5.78 & --- & 3.81 & --- & 5.30 & 7.20 & 3.32 \\ 
A478 & 0.0881 & 12.33 & 13.30 & 1.06 & 1.15 & 6.43 & --- & 6.82 & --- & 7.23 & 10.10 & 6.50 \\ 
A963 & 0.206 & 6.40 & 6.50 & 0.66 & 0.78 & 5.49 & 6.08 & 3.62 & 4.74 & 4.95 & 6.80 & 2.90 \\ 
A1689 & 0.1832 & 13.29 & 13.60 & 1.08 & 1.21 & 8.17 & --- & 8.82 & --- & 8.19 & 10.50 & 9.68 \\
A1795 & 0.0622 & 5.90 & 6.00 & 0.73 & 0.63 & 6.60 & --- & 4.82 & --- & 5.96 & 5.50 & 3.25 \\ 
A2029 & 0.0779 & 10.00 & 10.60 & 1.12 & 1.07 & 7.70 & --- & 8.62 & --- & 8.30 & 9.30 & 5.72 \\ 
A2261 & 0.224 & 9.97 & 12.00 & 0.93 & 1.65 & 6.23 & 6.10 & 5.79 & 10.07 & 6.41 & 14.40 & 8.33 \\ 
A2390 & 0.2329 & 17.20 & 17.30 & 1.54 & 1.73 & 8.89 & 10.28 & 13.69 & 17.78 & 10.35 & 15.20 & 11.72 \\ 
A3112 & 0.0759 & 3.84 & 4.40 & 0.40 & 0.47 & 5.02 & --- & 2.01 & --- & 3.67 & 4.10 & 1.27 \\ 
RX J0232 & 0.2836 & 12.53 & 13.30 & 1.07 & 1.45 & 6.41 & 10.06 & 6.86 & 14.59 & 6.95 & 12.70 & 6.07 \\ 
RX J0528 & 0.3839 & 10.55 & 11.60 & 1.11 & 1.52 & 6.04 & 7.80 & 6.70 & 11.86 & 6.88 & 13.30 & 8.33 \\ 
\\
A401 & 0.0743 & 5.82 & 6.80 & 1.02 & 1.16 & 7.26 & --- & 7.41 & --- & 7.65 & 10.10 & 5.86 \\ 
A520 & 0.203 & 7.11 & 8.40 & 1.13 & 1.37 & 7.74 & 7.23 & 8.75 & 9.91 & 8.11 & 11.90 & 6.99 \\ 
A665 & 0.1818 & 6.81 & 8.60 & 1.12 & 1.46 & 7.64 & --- & 8.56 & --- & 8.04 & 12.70 & 7.70 \\ 
A773 & 0.217 & 6.80 & 7.50 & 0.89 & 0.98 & 6.78 & 7.37 & 6.03 & 7.22 & 6.55 & 8.60 & 6.07 \\ 
A781 & 0.2984 & 4.75 & 6.00 & 0.76 & 0.90 & 5.72 & 5.10 & 4.35 & 4.59 & 5.35 & 7.90 & 5.08 \\ 
A1763 & 0.2279 & 8.00 & 10.50 & 1.14 & 1.94 & 6.55 & 6.32 & 7.47 & 12.26 & 7.37 & 17.00 & 9.04 \\ 
A1914 & 0.1712 & 10.73 & 11.30 & 1.07 & 1.21 & 8.26 & --- & 8.84 & --- & 8.19 & 10.60 & 7.06 \\ 
A2034 & 0.113 & 6.99 & 4.00 & 1.13 & 0.77 & 7.01 & --- & 7.92 & --- & 7.76 & 6.70 & 5.23 \\ 
A2163 & 0.203 & 23.86 & 28.70 & 3.17 & 4.40 & 13.40 & 12.27 & 42.48 & 53.99 & 19.68 & 38.50 & 32.13 \\ 
A2218 & 0.171 & 5.41 & 5.10 & 0.73 & 0.82 & 5.23 & --- & 3.82 & --- & 5.13 & 7.20 & 5.44 \\ 
A2219 & 0.2281 &14.94 & 15.50 & 1.74 & 2.16 & 9.37 & 10.90 & 16.30 & 23.54 & 11.44 & 18.90 & 16.53 \\ 
A2255 & 0.0809 & 2.47 & 2.90 & 0.59 & 0.67 & 5.79 & --- & 3.42 & --- & 4.91 & 5.90 & 3.67 \\ 
A2256 & 0.0581 & 3.92 & 4.60 & 0.78 & 0.82 & 6.40 & --- & 4.99 & --- & 6.11 & 7.20 & 5.01 \\ 
A3266 & 0.0602 & 4.22 & 4.90 & 0.96 & 1.06 & 7.46 & --- & 7.16 & --- & 7.51 & 9.20 & 6.36 \\ 
A3558 & 0.048 & 3.54 & 3.70 & 0.67 & 0.73 & 4.78 & --- & 3.20 & --- & 4.77 & 6.40 & 2.97 \\ 
A3921 & 0.094 & 1.28 & 3.10 & 0.29 & 0.62 & 5.03 & --- & 1.46 & --- & 3.03 & 5.40 & 2.33 \\ 
RX J0043 & 0.2924 & 8.26 & 7.70 & 0.88 & 0.92 & 5.82 & 7.59 & 5.12 & 6.98 & 5.88 & 8.10 & 9.89 
\enddata
\label{tab:pm_sample}
\tablecomments{Units are $10^{44}$ ergs/s for $L_X$, $10^{14}\ \msun$ for mass, $\keV$ for temperature, and $10^{14}\ \msun\keV$ for
$Y_X$ and $C^{-1}D_A^2\Ysz$.  Precise definitions of all quantities are specified in the text.    
The list is divided by the blank row into cool core (upper) and no cool core (lower) systems. }
\end{deluxetable*}




\section{B: Cluster Profiles and Aperture Corrections}
\label{app:aperture}

We consider the mean radial profile of three quantities:  gas density, $\rhogas (R)$, electron temperature $T(R)$, and 
pressure $P(R)$.  We use $x=R/\Rf$.  For $\rhogas$, we use the \citet{prattarnaud02} parameterization as reported by \citet{piffarettietal11},
\be
\rhogas(x) \propto \left( \frac{x}{x_c} \right)^{-\alpha}\left( 1+ \frac{x}{x_c} \right)^{-3\beta/2+\alpha/2.0}
\ee
with $x_c=0.303$, $\alpha=0.525$, and $\beta=0.768$. The temperature profile is parameterized by \citet{vikhlininetal06} as
\be
T(x) \propto \frac{ 0.45+ r }{1+r} \frac{1}{1+(x/0.6)^2}
\ee
where $r=(x/0.045)^{1.9}$.  Finally, \citet{arnaudetal10} parameterize the pressure profile as
\be
P(x) \propto \frac{1}{(cx)^\gamma \left[1+(cx)^\alpha\right]^{(\beta-\gamma)/\alpha}}
\ee
with $c=1.177$, $\gamma=0.3081$, $\alpha=1.0510$, and $\beta=5.4905$.  

We use each of these profiles to apply aperture corrections for observables when necessary.  Specifically,
given an observable $X$ evaluated at $R=\Rf$, we investigate how a bias $b$ in the mass estimate
propagates into cluster observables due to the choice of aperture.  To do so, we use the above models
to estimate the logarithmic slope $\epsilon = d\ln X/d\ln R$, so that $X\propto R^\epsilon$ at $R=\Rf$.
Since $\Rf\propto M^{1/3}$, we see that bias $b$ in the mass will result in a bias $b^{\epsilon/3}$
in the observable of interest.

We first consider luminosity.  Before we evaluate the logarithmic slope for $\Lx$, we first use our model to
learn how to rescale total luminosities to luminosities within a circular aperture $\Rf$. 
As detailed in \citet{piffarettietal11}, the total soft X-ray band luminosity is relatively insensitive to temperature, so that $\Lx \propto \int dV \rhogas^2$, from which one finds $\Lx(\Rf)/\Lx(\infty) = 0.91$, where $\Lx(\Rf)$ is the luminosity within a circular aperture $R=\Rf$ (so that the integration region is cylindrical).  In this calculation, we have truncated the integral at $R=5\Rf$, a practice we adopt throughout.  Our results are very weakly dependent on this choice.  In practice, however, this correction appears to
overestimate the luminosity contribution of the cluster outskirts, so we use a value of $0.96$ recovered from explicit
reanalysis of clusters by A. Mantz (private communication).

Using the \citet{piffarettietal11} model, we calculate the logarithmic
slope $\epsilon = d\ln L/d\ln R$ at $R=\Rf$, finding $\epsilon=0.16$.  Thus, $\Lx \propto b^{0.05}$ where $b$
is the bias in the mass.  This is an exceedingly weak dependence: setting $b=1.3$, we find that the corresponding change
in luminosity is just over $1\%$.  Similarly, 
we have $\Mgas \propto \int dV \rhogas$ where the integral is now over a spherical region.  Our adopted model
results in a logarithmic slope at $R=\Rf$ of $\epsilon=1.20$, or $\Mgas \propto b^{0.40}$.

We next turn to temperature.   Following \citet{mazzottaetal04}, we assume that the spectroscopic temperature $\Tx$ 
is given by $\int dV wT /\int dV w$ where $w=\rhogas^2 T^{-3/4}$ \citep[see also][]{vikhlinin06}.  We find
$\Tx \propto b^{-0.15}$.   This is a sufficiently weak dependence that it can be ignored for our purposes, and reflects the
fact that the temperature profile is very flat at around $R\approx 0.15\Rf$. 
Finally, because $\Yx$ is defined as the product of $M_{gas}$ and $\Tx$, the sensitivity
to bias is just the product of the bias scalings for $M_{gas}$ and $\Tx$ individually.   Thus, one has
$\Yx\propto b^{0.40}\times b^{-0.15} = b^{0.25}$.  

Finally, turning towards the integrated pressure $Y \propto \int dV P$, using the \citet{arnaudetal10} profile,
we find $Y \propto R^{0.82}$ at $R=\Rf$.  Consequently, a bias $b$ in the mass results in a bias 
$b^{0.27}$ in the integrated $Y$.


\section{C:  Interpreting the Evolution in the (P11-LS)--M10 Comparison}
\label{app:m_evol}

There appears to be redshift evolution in the (P11-LS)--M10 mass ratio.  Here, we 
quantify the significance of this evolution, and attempt to determine which of the
two data sets might be systematically different at high redshift.   We do not arrive at
any firm conclusions, but do find some hints of possible systematics that require further 
investigation.

\subsection{Significance of the Observed Evolution}
\label{sec:significance}

We quantify the evolution in the (P11-LS)--M10 comparison by fitting the redshift behavior of the mass differences to  
\be
\Delta \ln M (z) = a+\gamma \ln \left( \frac{1+z}{1+0.15} \right).
\label{eq:evoleq}
\ee
As per the discussion in section \ref{sec:errors}, we use uniform weighting in our fits, and rely on bootstrap resampling
for all statistical inferences.  Cluster A2034 is not included in the fits.  We find $\gamma = -1.90 \pm 0.44$,
which differs from $\gamma=0$ at $4.3\sigma$.  Thus, the observed evolution is statistically significant.
However, part of this evolution is not intrinsic redshift evolution, but rather induced evolution: because of volume
effects, the (P11-LS)--M10 cluster sample at high redshift is more massive than the low redshift sample.  If the systematic
mass offset scales with mass, then this selection effect would induce apparent redshift evolution.
To test for this possibility, we fit a power-law model with lognormal scatter to the mass offset $\Delta \ln M$ as a function of $\Ysz$.
We then randomly assign a new $\Delta \ln M$ to each cluster based on its own $\Ysz$ value, irrespective of redshift,
and measure the resulting redshift evolution.  The whole procedure is iterated $10^4$ times to evaluate the mean
induced evolution and the corresponding uncertainty.

\begin{figure}
\begin{center}
\hspace{-0.2 in} \scalebox{0.6}{\plotone{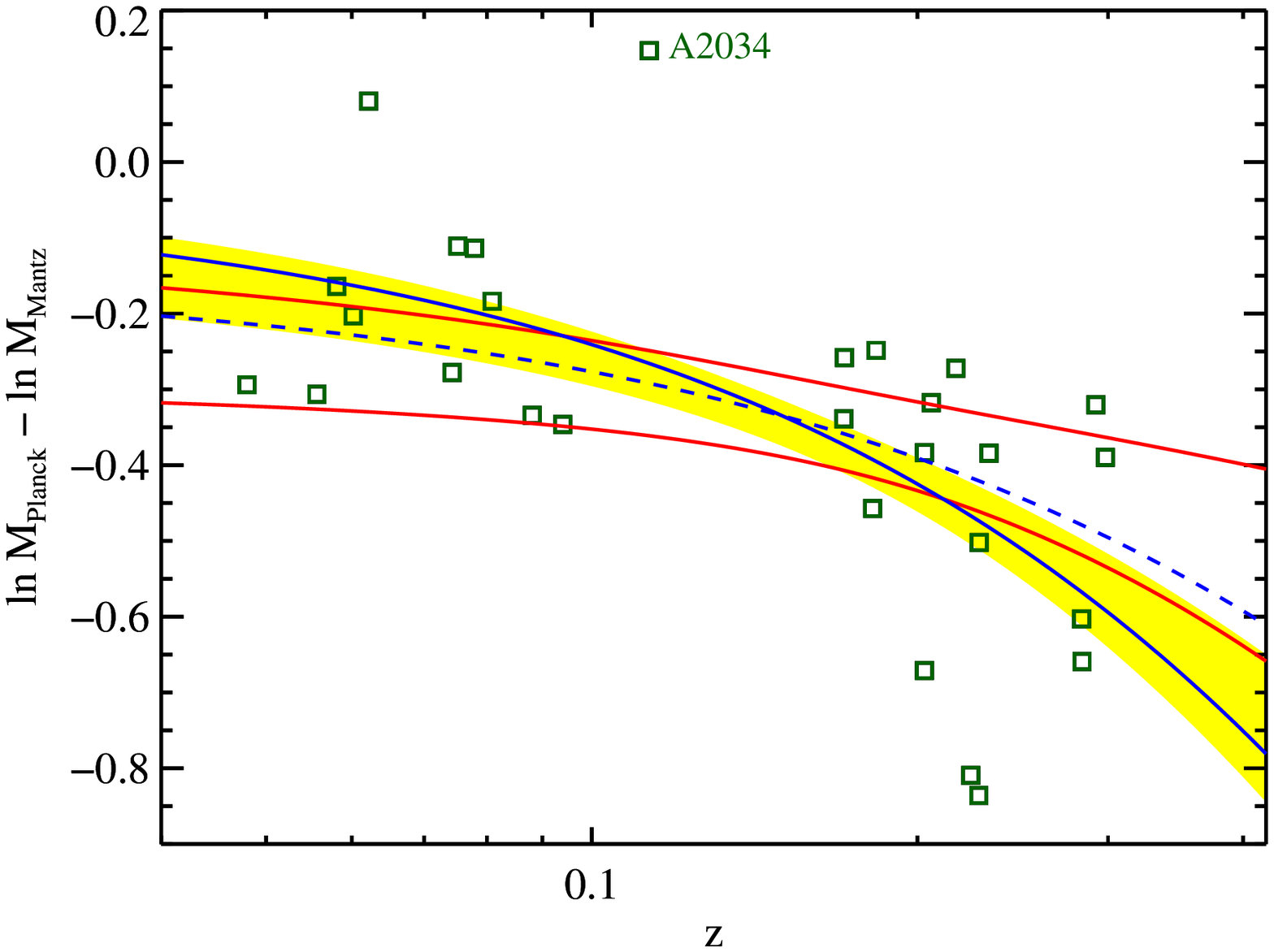}}
\hspace{-0.1 in}\scalebox{0.6}{\plotone{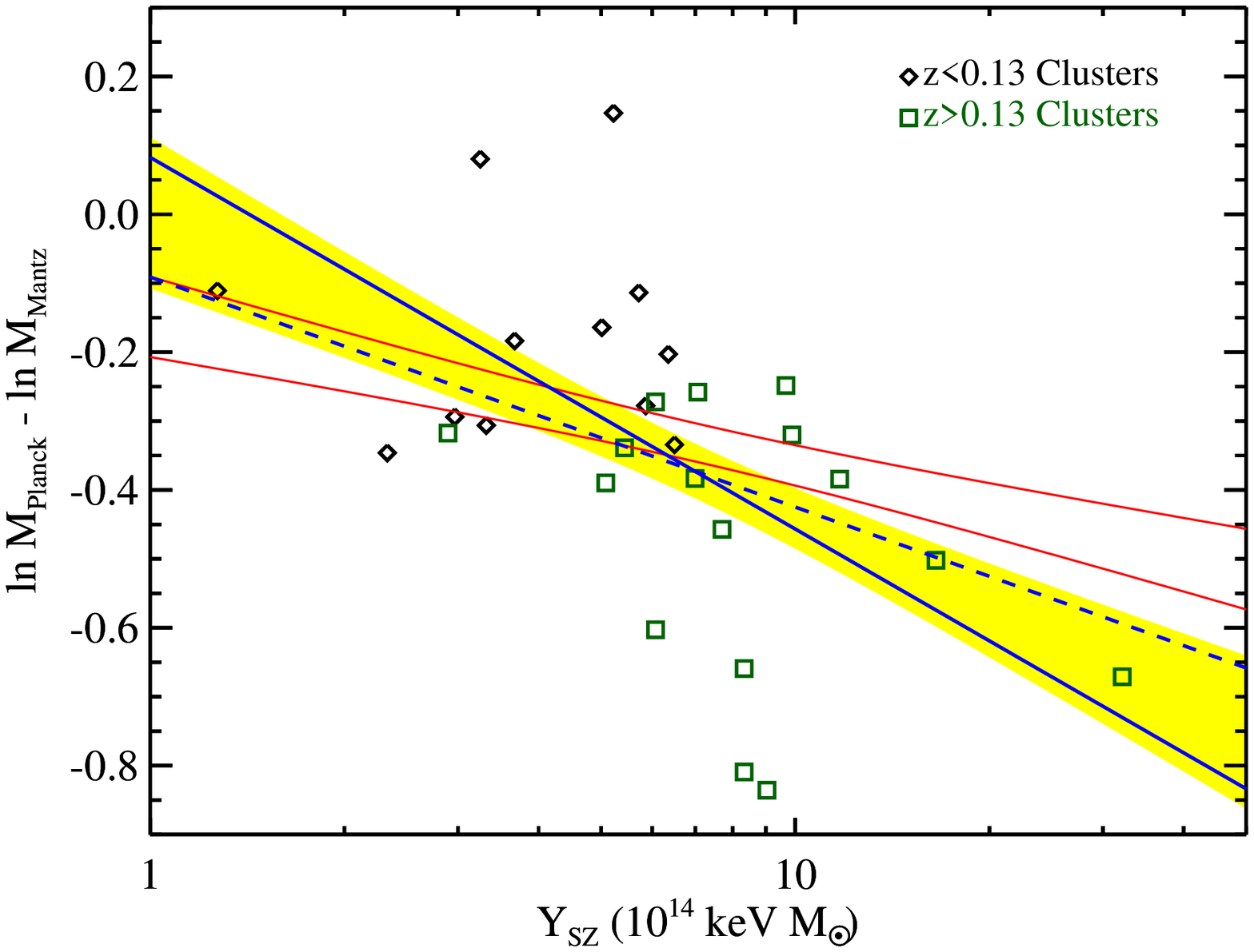}}
\caption{{\it Left panel: } (P11-LS)--M10 mass offset, $\Delta \ln M$, as a function of redshift.   
The yellow band shows the $1\sigma$ confidence band of the best fit
scaling relation, while the red band shows the redshift evolution induced by selection effects in
a model where the $\Ysz/\Yx$ ratio scales with $\Ysz$ as observed in the right panel.
{\it Right panel: }The (P11-LS)--M10 offset $\Delta \ln M$, as a function of the SZ signal $\Ysz$.   The yellow band shows our best
fit model.  The clusters are split into low and high redshift clusters for illustrative purposes only: the fit
uses all galaxy clusters (after outlier rejection).  The red solid lines shows the scaling induced
by selection effects alone plus intrinsic redshift evolution as observed in the left panel.
In both panels, the solid and dashed blue lines are the predicted scalings for the \citet{prattetal09} and \citet{linetal11}
$\fgas$ models.
}  
\label{fig:evol1}
\end{center}
\end{figure}

The left panel in Figure \ref{fig:evol1} shows the $\Delta \ln M$ vs $z$ for the (P11-LS)--M10 comparison.  The yellow band is the best
is the $68\%$ confidence region for a power-law model for the evolution, i.e. Equation \ref{eq:evoleq}.
The induced evolution due to scaling
of $\Delta \ln M$ with mass is shown with red error bands.  The two are consistent, but there is some suggestion that there
is intrinsic evolution in $\Delta \ln M$ with redshift.
We have also carried the reverse calculation: i.e. used our best fit redshift evolution to make synthetic realizations of
our data so as to determine the induced $\Ysz$ evolution.  The corresponding data, observed evolution (yellow band), 
and redshift-induced evolution (red band) are shown in the right panel of Figure \ref{fig:evol1}.


In short, it is apparent from Figure \ref{fig:evol1} that the mass offset in the (P11-LS)--M10 comparison depends
on cluster mass, or cluster redshift, or, more likely, both.  We now consider two alternate hypothesis:
1- the masses in P11-LS are correct, and the evolution in $\Delta \ln M$ is entirely due to a systematic effect in M10.
2- the masses in M10 are correct, and the evoultion in $\Delta \ln M$ is entirely due to a systematic effect in P11-LS.


\subsection{Are the \citet{mantzetal10b} Masses Systematically Evolving?}
\label{sec:evolution_model}


The foundation of our investigation is a critical observational fact:
all groups are in excellent agreement when it comes to estimating the $\Mgas(R)$ profile.  
Since the masses in M10 are based on $\Mgas$, it follows that any bias must ultimately
be sourced by the constant $\fgas=f_0$ model adopted in M10.
Assume then that some other $\fgas(M,z)$ is correct.  Assuming $\Mgas(R)\propto R^\gamma$ at
$R=\Rf$, one has $\Mgas(R)=M\fgas(M)(R/\Rf)^\gamma$.  Given $f_0$, M10 assigns a mass by the condition
$\Mgas(R_0) = f_0 M = f_0 C R_0^3$ where $C=(4\pi/3)\Delta_c \rho_c$.  Setting the two expressions
for $\Mgas(R_0)$ equal to each other, we can solve for $R_0/Rf$  in terms of $\fgas/f_0$.  Converting to
mass, we arrive at
\be
b_{M10} = \frac{M_{M10}}{M} = \left( \frac{\fgas(M,z)}{f_0} \right)^{1.67} \label{eq:bias1}
\ee
where we used $\gamma=1.2$ as per Appendix \ref{app:aperture}.

If one sets $\fgas(M,z)$ to a model consistent with the P11-LS data, one will recover the observed
evolution in the (P11-LS)--M10, as this is simply a restatement of the fact that the two works recover
the same $\Mgas(R)$ profile.  We demonstrate this explicitly.
The left panel in Figure \ref{fig:evol1} shows the
predicted offset from the \citet{prattetal09} $\fgas$ model, where the redshift dependence
is included assuming an $\fgas$ evolution model $\fgas \propto (M/M_*(z))^\alpha$ where
$\alpha$ is the slope of the $\fgas$--$M$ relation.  In \citet{linetal11},
the observed evolution is $(1+z)^{0.41}$, which corresponds to $\alpha=0.13$.
The z-dependence for $\alpha=0.21$ \citep[][]{prattetal09} is therefore $(1+z)^{0.41\times 0.21/0.13}$,
or $\fgas\propto M^{0.21}(1+z)^{0.66}$.  The total evolution predicted by this model
is the blue solid line, in excellent agreement with the observed (P11-LS)--M10 offset.  Also shown
as a blue dashed line is the \citet{linetal11} model, which under-predicts the observed
evolution because of the shallower $\fgas$--$M$ slope.

The take-home message is that the M10 masses can only be biased if the constant $\fgas$ model
is incorrect.  The question then becomes, can we be confident that $\fgas$ scales with mass and/or redshift?  
From a theoretical point of view,
the expectation from hydrodynamic simulations with no feedback or gas cooling is a weakly evolving gas 
fraction with mass or redshift \citep[see e.g.][]{staneketal10}.  Some simulations that include baryonic feedback 
exhibit constant $\fgas$ \citep[e.g.][]{fabjanetal11}, but others do not \citep{nagaietal07b, vikhlininetal09}.  
In short, theory does not yet give us an unequivocal answer as to whether $\fgas$ scales with
mass and/or redshift.

Observationally, things are also not clear cut.   \citet{allenetal08}
finds that $\fgas$ does not scale with cluster temperature: $\fgas \propto T^{0.02\pm 0.06}$, suggesting
no mass-scaling.  We note, however, that the
fact that these are not core-excised temperature, cool-core clusters complicates the interpretation of this measurement.
Moreover, \citet{allenetal08} worked at $R_{2500}$ rather than $\Rf$.  
If we ignore these difficulties, and assume a self-similar
scaling $T\propto M^{2/3}$, we find that 
$\fgas \propto M^{0.00\pm 0.04}$ \citep[though the slope of the $\Tx$--$M$ relation is itself the source of active investigation, see e.g.][]{mantzallen11},
at a pivot point $M_p\approx 10^{15}\ \msun$.  Other representative values for the slope of the $\fgas$--$M$ relation 
from the literature are $\alpha=0.13 \pm 0.02$ \citep{vikhlininetal09}, $\alpha=0.21 \pm 0.03$ \citep{prattetal09}, 
$\alpha=0.09 \pm 0.03$ \citet{sunetal09}, $\alpha=0.13\pm 0.03$ \citep{linetal11}.  The range of values in the
literature tend to span the range $\alpha \in [0.1,0.2]$, which is in mild to modest tension with the \citet{allenetal08} value.
These differences can be reconciled if the $\fgas$--$M$ flattens at  high masses, as the \citet{allenetal08} has a higher
pivot point that the remaining studies.  If this is not the case, then there are clearly systematics that remain to be 
fully addressed.  Some examples of possible systematics are the use of X-ray temperatures that are not core-excised in \citet{allenetal08},
or parameterization systematics as argued by \citet{mantzallen11}.

Things are even less clear when it comes to the redshift dependence of $\fgas$.  In both \citet{allenetal08} and
\citet{mantzetal10b}, it is assumed that there is no intrinsic redshift evolution in $\fgas$, and any redshift evolution is interpreted
as evidence of errors in the underlying cosmological model used to interpret the observational data.
\citet{vikhlininetal09} relied on numerical simulations from \citet{nagaietal07b} to inform their $\fgas$ model,
which is in turn used to estimate $\Mf$ from the $\Mgas$ data, so again the redshift evolution of $\fgas$
was not directly probed.  \citep{linetal11} measure $\fgas\propto (1+z)^{0.41\pm 0.14}$, but they had to assume
self-similar evolution for the $M$--$\Yx$ relation, and even this is only $3\sigma$ away from a no-evolution
model.  Overall, the question of whether or not $\fgas$ evolves with redshift remains an open question.


\subsection{Are the \citet{planck11_local} Masses Systematically Evolving?}
\label{sec:evolution_test}

Having considered the possibility that the evolution in the (P11-LS)--M10 comparison is due
to systematic evolution in the \citet{mantzetal10b} masses, we now consider the converse
scenario; we assume that the \citet{mantzetal10b} masses are correct, and the observed
evolution is entirely due to systematic evolution in the P11-LS masses, and test any
observational implications of this scenario.

Once again, the foundation of our exploration is the observation that the $\Mgas$ profile --- and therefore
the $\rho_{gas}(R)$ profile --- is
robustly and correctly estimated by all groups.  It follows then that any evolution in the P11-LS masses
must be due to systematic trends in the X-ray temperature.
We can test this hypothesis by constructing an SZ temperature $T_{SZ}=\Ysz/\Mgas$ that can be compared
to the X-ray temperature $\Tx$.  Any evolution in this ratio --- which is none other than the $\Ysz/\Yx$
ratio --- would signal a systematic trend in $\Tx$ with cluster redshift.   Moreover, to explain
the evolution in the (P11-LS)--M10 comparison, the evolution in
the $\Ysz/\Yx$ ratio of \citet{planck11_local} should evolve from low to high values as one moves
higher in redshift.


\begin{figure}
\begin{center}
\hspace{-0.2in} \scalebox{0.6}{\plotone{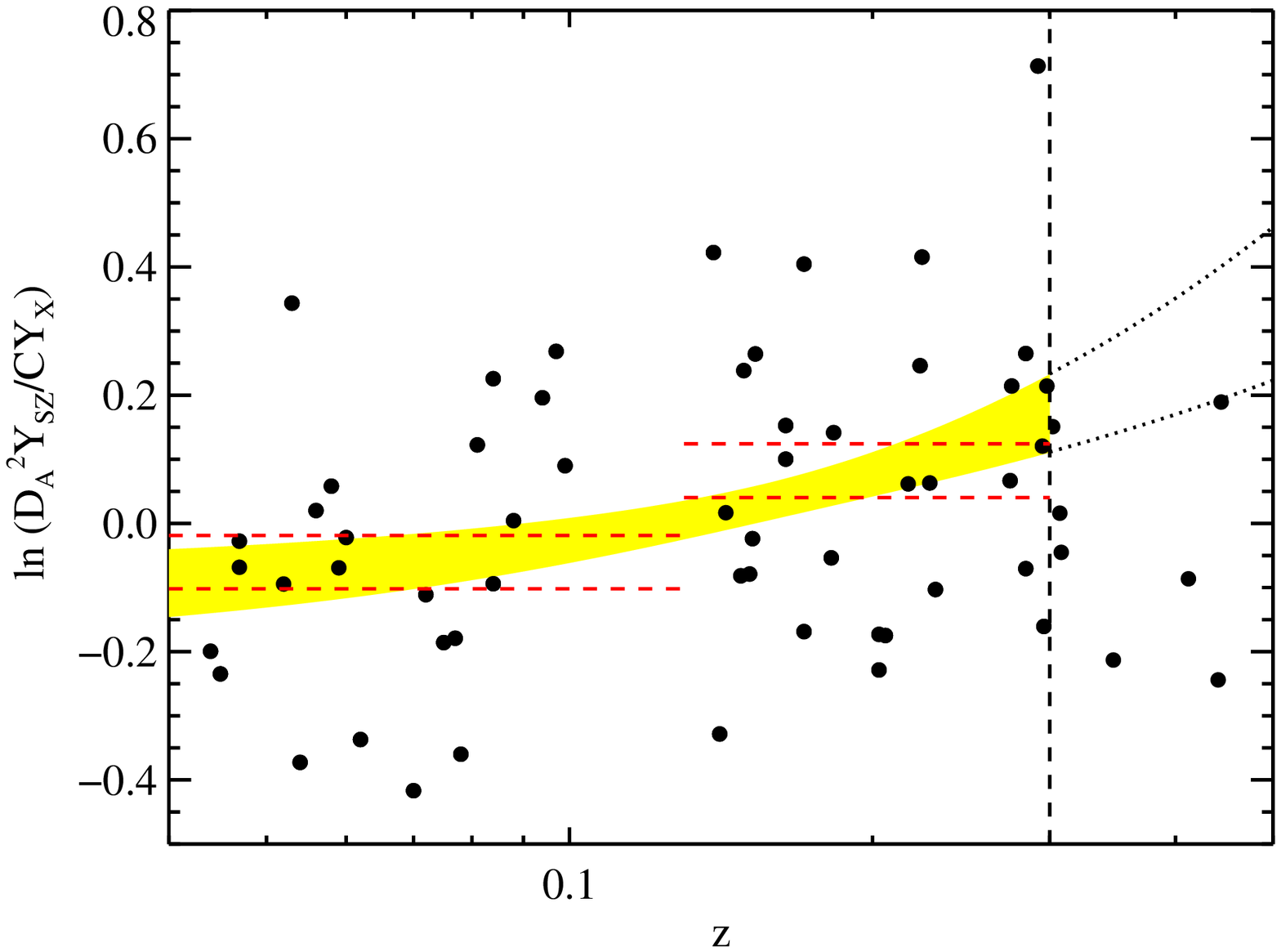}}
\hspace{-0.1in} \scalebox{0.6}{\plotone{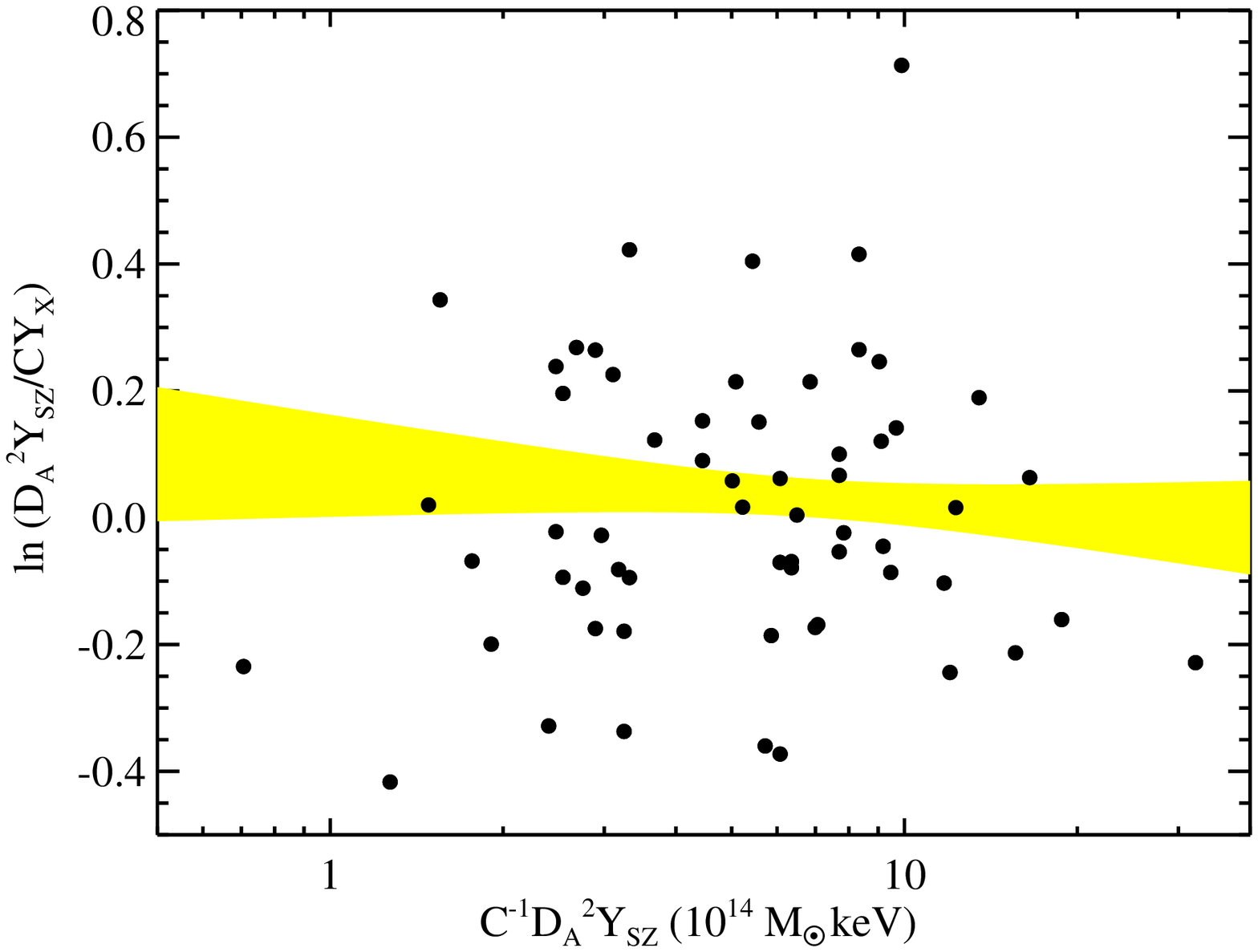}}
\caption{The $\Ysz/\Yx$ ratio as a function of redshift (left panel) and $\Ysz$ signal (right panel) 
in the \citet{planck11_local} data.  The best
fit power-law model $\Ysz/\Yx\propto (1+z)^\gamma$ results in $\gamma=1.19 \pm 0.43$, which
differs from zero at the $2.7\sigma$ level.  The hypothesis $\gamma=0$ can be ruled out at
$99.6\%$ confidence.  Both of these results use only $z\leq 0.3$ clusters (dashed vertical line), as per the rest of the paper.
The red-dashed horizontal lines denote the $68\%$ confidence interval for the $\Ysz/\Yx$ ratio for low
and high redshift clusters, as shown.
}  
\label{fig:ysz_yx}
\end{center}
\end{figure}


The top panel in Figure \ref{fig:ysz_yx} shows the evolution in the $\Ysz/\Yx$ ratio using the 
\citet{planck11_local} data.  For the figure, we
have multiplied the $\Yx$ value reported in \citet{planck11_local} by a factor of 0.95 so that it matches 
the definition adopted throughout the paper (i.e. $\Tx$ is to be measured within $[0.15,1]\Rf$, see
section \ref{sec:tx} for details).  To quantify evolution, we fit a power-law to all clusters  with $z\leq 0.3$,
except for A2034, which has an incorrect redshift.  
The redshift cut is also motivated by the fact that all the cross-comparisons we have performed
have all focused on $z\leq 0.3$ clusters.  We note, however, that our conclusions
are sensitive to this cut, a point that we will return to shortly.

A power-law model $\Ysz/\Yx \propto (1+z)^\gamma$ fit to the data 
results in $\gamma=1.19\pm 0.43$, which deviates from the no evolution expectation 
at the $2.7\sigma$ level.  The $68\%$ confidence band defined by our best fit
model is shown in Figure \ref{fig:ysz_yx}.  Note that
while the Figure extends to $z>0.3$, the fit only includes $z\leq 0.3$ clusters. The extrapolation
of the fit to higher redshifts is shown by the dotted lines.
We have also estimated the probability at which the hypothesis $\gamma=0$ can be
ruled out through a direct Monte Carlo experiment.  First, we compute the average
$\Ysz/\Yx$ ratio over all clusters, and estimate the corresponding scatter.  
We use a log-normal model to assign $\Ysz/\Yx$ to every cluster independently
of its redshift, and measure the corresponding redshift evolution as we have done
in the data.  The process is repeated $10^5$ times.  The confidence at which
we can rule out the $\gamma=0$ hypothesis is the fraction of times for which 
the slope $\gamma$ in our synthetic data falls below the observed value.   
The $\gamma=0$ hypothesis is ruled out with $99.7\%$ confidence.

As noted earlier, however, these
results are sensitive to whether we cut the \citet{planck11_local} cluster
sample at $z=0.3$ or not.   Removing the $z\leq 0.3$
redshift cut adds a few systems with lower $\Ysz/\Yx$ ratios that drive 
the redshift evolution lower.  Indeed, visual inspection of Figure \ref{fig:ysz_yx}
gives the distinct impression that there are two distinct cluster populations,
split at $z=0.13$.
 We have computed the mean $\Ysz/\Yx$ ratio for $z\leq 0.13$ and $z\in[0.13,0.3]$
galaxy clusters.  
For the low and high redshift systems we find $\ln (\Ysz/\Yx) = -0.06\pm 0.04$
and $\ln(\Ysz/Y-X)=0.08\pm 0.04$.   These are shown as the red-dashed lines in the
left panel of Figure \ref{fig:ysz_yx}.
The two values differ at the $2.5\sigma$ level: interesting, but not
high enough significance to draw robust conclusions. 

There are two things worth noting about this result.  First: if this difference is real, it is {\it not}
due to selection effects.   Indeed, P11-LS have already demonstrated that --- because of the tight scatter between
$\Ysz$ and $Y_X$ --- selection effects have a negligible impact on the recovered scaling relation. 
One might expect that higher S/N clusters --- which preferentially
reside at lower redshift because of the Planck beam --- would have a higher $\Ysz/Y_X$ ratio, which could
in principle explain the observed evolution.   However, this does not appear to be sufficient
to explain the apparent offset.  A linear fit to the $\Ysz/Y_X$ ratio as a function of S/N result in
$\Ysz/Y_X \propto (S/N)^{-0.09}$.  The median S/N for $z\leq 0.13$ and $z\in[0.13,0.3]$ is
$S/N=11.3$ and $S/N=9.2$ respectively.  The induced shift in the $\Ysz/Y_X$ ratio is
$\Delta \ln (\Ysz/Y_X)_{induced} = 0.09\ln(11.3/9.2) = 0.02$, which is to be compared to the observed
shift of $\Delta \ln (\Ysz/Y_X)=0.14$.  Evidently, this type of selection effect cannot account for the
observed difference between the two cluster populations.

In addition, it is worth emphasizing that 
the $\Ysz/\Yx$ ratio is larger than unity for high redshift systems.  There is now ample
empirical evidence that clusters are not isothermal, but rather 
have falling temperature 
profiles \citep[e.g.][]{vikhlininetal06,prattetal07,sunetal09,arnaudetal10,morettietal11}.
Because of the $\rho^2$ vs $\rho$ weighting of the X-ray and SZ signals, the average
X-ray temperature is dominated by a region interior to that dominating the SZ signal,
which, in conjunction with a falling temperature profile, predicts $\Ysz/Y_X \leq 1$.
The fact that the P11-LS $\Ysz/Y_X$ ratio at $z\in[0.13,0.3]$ violates this inequality
is therefore problematic.

For completeness, the right panel of Figure \ref{fig:ysz_yx} shows the dependence
of the $\Ysz/\Yx$ ratio on $\Ysz$, which acts as a mass proxy.  
We find $(\Ysz/\Yx)\propto \Ysz^{-0.03\pm0.04}$, which is consistent 
with zero.  Note this implies that any induced evolution due to mass selection
effects must also be minimal, in agreement with the P11-LS results on the impact
of selection effects on the $\Ysz$--$Y_X$ relation.


\end{document}